\documentclass[submission, Phys]{SciPost}

\hypersetup{
    colorlinks,
    linkcolor={red!50!black},
    citecolor={blue!50!black},
    urlcolor={blue!80!black}
}

\usepackage[bitstream-charter]{mathdesign}
\DeclareSymbolFont{usualmathcal}{OMS}{cmsy}{m}{n}
\DeclareSymbolFontAlphabet{\mathcal}{usualmathcal}

\fancypagestyle{SPstyle}{
\fancyhf{}
\lhead{\colorbox{scipostblue}{\bf \color{white} ~SciPost Physics }}
\rhead{{\bf \color{scipostdeepblue} ~Submission }}

\fancyfoot[C]{\textbf{\thepage}}
}

\usepackage{graphicx} 
\usepackage{amsmath}
\usepackage{xcolor}
\usepackage{enumitem}

\begin{document}
\pagestyle{SPstyle}

\begin{center}{\Large \textbf{\color{scipostdeepblue}{
How Molecular Motors' Interaction\\ Shapes Flagellar Beat and Its Fluctuations
}}}\end{center}


\begin{center}\textbf{
F. Fanelli\textsuperscript{1,2$\star$},
A. Puglisi\textsuperscript{3,1,4$\dagger$},
}\end{center}

\begin{center}
{\bf 1} Sapienza University of Rome, Physics Department, P.le A. Moro 2, Rome, Italy
\\
{\bf 2} Sorbonne Université, INSERM, Institut Pierre Louis d’Epidémiologie et de Santé Publique, IPLESP, Paris, France
\\
{\bf 3} Istituto dei Sistemi Complessi - Consiglio Nazionale delle Ricerche, P.le A. Moro 2, Rome Italy
\\
{\bf 4} INFN, Sezione Roma2, Via della Ricerca Scientifica 1, I-00133, Rome, Italy
\\[\baselineskip]
$\star$ \href{mailto:federico.fanelli2001@gmail.com}{\small federico.fanelli2001@gmail.com}\,,\quad
$\dagger$ \href{mailto:andrea.puglisi@cnr.it}{\small andrea.puglisi@cnr.it}
\end{center}

\section*{\color{scipostdeepblue}{Abstract}}
\textbf{\boldmath{%
{\color{red}The stochastic dynamics of flagellar beating for micro-swimmers, such as flagellated cells, sperms and microalgae, is widely thought to include a feedback mechanism between flagellar shape and the rate of activation/de-activation of the $N \gg 1$ driving molecular motors.} In the context of the so-called rigid filament models, where the axoneme is described by a single degree of freedom $X(t)$, we investigate the effect of direct coupling between the activity dynamics of adjacent motors, parametrized by $K \ge 0$. A functional Fokker-Planck equation for $X$ and the state of the $N$ motors is obtained. In the limit of small coupling $K \ll 1$, we derive a  system of equations governing the dynamics of the Fourier modes of the active motor density, obtaining estimates for several observables and the fluctuations' quality factor $Q$. For larger $K$ we resort to numerical simulations. The effect of introducing the coupling $K>0$ is to increase characteristic times and the beating period. Moreover for large $K$s the limit cycle becomes bi-stable, with abrupt avalanches of the motor dynamics. Increasing $K$ is similar to what observed in the case $K=0$ when the confining elastic force is strongly reduced. The quality factor of fluctuations has a non-monotonic behavior: it first increases with $K$, then decreases. This is accompanied by the reduction and eventual disappearance of regions where the fraction of activated motor is nor $0$ neither $1$.
}}

\vspace{\baselineskip}

\noindent\textcolor{white!90!black}{%
\fbox{\parbox{0.975\linewidth}{%
\textcolor{white!40!black}{\begin{tabular}{lr}%
  \begin{minipage}{0.6\textwidth}%
    {\small Copyright attribution to authors. \newline
    This work is a submission to SciPost Physics. \newline
    License information to appear upon publication. \newline
    Publication information to appear upon publication.}
  \end{minipage} & \begin{minipage}{0.4\textwidth}
    {\small Received Date \newline Accepted Date \newline Published Date}%
  \end{minipage}
\end{tabular}}
}}
}


\vspace{10pt}
\noindent\rule{\textwidth}{1pt}
\tableofcontents
\noindent\rule{\textwidth}{1pt}
\vspace{10pt}

\section{Introduction}

The dynamics of active biological structures, such as motorized organelles, cilia and flagella, results from a complex interplay of elasticity, hydrodynamics and active driving~\cite{gilpin2020multiscale}. The last ingredient, usually in the form of several molecular motors converting chemical energy into mechanical displacement, deserves investigation from many different perspectives: biochemistry, thermodynamics, non-equilibrium statistical physics~\cite{cicuta2020use}. The energy injection rate of molecular motors is a stochastic process, only partially understood, and its impact on the mechano-hydrodynamics of mesoscopic objects is still far from a clear picture~\cite{sanchez2011cilia,ling2018instability,cecconi2025active}. Fluctuations of the observed oscillations in these systems can provide a powerful magnifying lens, disclosing internal details that are not yet directly accessible in vivo~\cite{Ma2014,ferretta2024thermal,kotz2025motor}. However this magnifying lens needs theory to be fully exploited,  exactly as in the primordial example of this kind, Brownian motion: Einstein provided an essential model, Perrin exploited it in experiments, and the properties of atoms were disclosed to the eyes of scientists, previously blind~\cite{smith2020brownian}. Even more than a century later, now in the biological context, good models and their analysis are still crucial to fill the gap between microscopic mechanisms, occurring at the nano-scale, and mesoscopic observation, at the micro-scale~\cite{Ma2014,Maggi2022,Sharma}. The presence of energy conversion from a chemical storage into directed motion, and finally dissipation, imposes a time-arrow in these systems, breaks time-reversibility and implies the impossibility to apply the principles of equilibrium statistical physics~\cite{barato2015thermodynamic,horowitz2020thermodynamic,Maggi2022}. 

A neat example of this problem is provided by flagellar beating, such as that occurring in cilia and flagella as appendices of eukaryotic cells, e.g. in sperms, certain micro-algae, or in several epithelia~\cite{gilpin2020multiscale}. In all such examples, the internal structure of a flagellum is the same and takes the name of axoneme. An extreme simplification of this structure reduces it to 9 doublets (excluding the central one which is not fundamental for the rest of the discussion) of microtubules (MTs), running parallel to each other all along the length of the flagellum. Each doublet is decorated by thousands of dyneins, which are proteins (ATP-ases) acting as molecular motors by keeping an arm anchored to the first MT and attaching/detaching from the second one. This attaching/detaching process follows a stochastic dynamics that consumes ATP and induces local sliding between the MTs. Several mechanical constraints in the axoneme convert such a relative sliding  into local bending. How hundreds of thousands of stochastic bending strokes coordinate to generate a smooth travelling wave is a fascinating and debated matter~\cite{riedel2007molecular}. {\color{red}Many models assume a feedback mechanism is needed: active motors induce local bending, bending makes (locally) the motors de-activate, but then the bending relaxes and the motors activate again~\cite{brokaw1975molecular,Julicher1997}.
Other models support the possibility that oscillations can arise  from mechanical instability even in the absence of such feedback~\cite{Bayly2016,Bigoni2023,DeCanio2017,Woodhams2022}.} Feedback loops, which take place in space and time, can be realised according to several scenarios. For instance the feedback response can be sensitive to different local observables, such as relative parallel displacement between the MTs~\cite{brokaw2005computer}, curvature~\cite{machin1958wave,brokaw2002computer} or its time-derivative~\cite{sartori2016dynamic}, or normal/tangential forces/deformations~\cite{lindemann2002geometric}. Experiments have provided some discrimination, but not unique across different biological objects~\cite{riedel2007molecular}. Much less studied is how fluctuations are affected by these feedback mechanisms. Theory and experiments, according to our view, are still in their infancy for this particular problem~\cite{Ma2014,Maggi2022}. A fundamental model taking into account the finiteness of the number of motors, and therefore their fluctuations,  has been used only a few times to analyse experimental data: in the following we call it J\"ulicher-Prost (JP) model~\cite{Julicher1997}. 

A modification of the JP model has been recently proposed to understand new experimental observations: the modification consists in adding a direct coupling between adjacent motors, inducing stronger correlations and consequently more complex fluctuations in the  spatio-temporal beating pattern~\cite{Maggi2022}. The impact of noise correlation (analogous to motor coordination) for fluctuations of the beating limit cycle has been studied also in~\cite{costantini2024thermodynamic,ferretta2024thermal,Sharma}. In the rest of the paper we call this model JPK i.e. JP with coupling constant $K$ (when $K \to 0$ the original JP model is reproduced). 

We mention that another class of models exists, the so-called crossbridge or ‘‘power-stroke’’ models~\cite{huxley1957muscle,brokaw1975molecular,vilfan1999force}, where the motors are described as flexible springs, with their heads binding to specific  sites of the filament and remaining stuck until unbinding. The motors switch between conformational states. The dynamic instability is associated to a variation of the unbinding rate with spring tension, in particular the rate should be increasing with the spring stretching. An interesting connection between the crossbridge model and the JP model - involving also nano-tribology, has been made in~\cite{guerin2010dynamic}.

In the present paper we  investigate, numerically and analytically, the JPK model, with the aim of illustrating the main macroscopic modifications induced by the direct nearest-neighbor coupling with respect to the original JP model. These modifications can be summarised as: 1) for larger and larger $K$ new dynamical interactions between spatial modes appear, i.e. more and more time-scales are relevant, making relaxations and cycle periods much longer; 2) for small $K>0$, analytical predictions can be obtained for the evolution of the motor density modes, for the quality factor of fluctuations and approximated estimates of the critical condition for the appearance of limit cycle and its critical frequency; for larger $K$, numerical simulations show a change in the shape of the beating limit cycle, with a much less uniform dynamics or - said differently - more rapid - bi-stable or avalanche-like - switching between extreme values: for very large $K$ in a given region motors are all-attached or all-detached; this is accompanied by a non-monotonous behavior of the quality factor of fluctuations: for small $K$ the precision increases but for larger $K$ it strongly decreases. Interestingly, from the point of view of the width and uniformity of the limit cycle, increasing $K$ is similar to decreasing elastic confinement in the original model at $K=0$. We mention that the model analysed here is slightly different from the one studied in~\cite{Maggi2022} and for this reason the results do not entirely overlap but are qualitatively consistent, see~\ref{rates} for details. 

In Section~\ref{model} we review the JP model and define the new JPK model under investigation. Then in Section~\ref{ffp} we re-derive the functional Fokker-Planck equation for the "slow" variables of the JPK model, that are the active motor density field $\rho(x,t)$ and the flagellar wave phase variable $X(t)$. In Section~\ref{analysis}, we consider the small $K$ limit, obtaining a system of equations for the Fourier modes of $\rho(x,t)$ coupled among themselves and to $X$. The last subsection of Section~\ref{analysis} employs some stronger approximation and truncations of the dynamics of the modes, leading to simpler interpretations and estimates for the transition to the limit cycle regime and its critical period. In Section~\ref{numerics} we discuss the results of numerical simulations of the model, which for small values of $K$ agree with the analytical theory, while for larger values depart from it and provide an interesting counterpart of recent experiments on sperm cells~\cite{Maggi2022}.  Finally, in Section~\ref{conclusions} we draw conclusions and propose perspectives. Two Appendices are present: the first discusses with details and examples a spatially homogeneous model, for pedagogical completeness; the second one presents all main analytical derivations of the main results presented in the paper.

\section{Models} \label{model}

In this Section we recall basic facts of the JP model ($K=0$) and we introduce the new JPK model investigated in the present work.

The general strategy of flagellar models with feedback is to decompose a flagellum into two main objects: a mechanical body and the motors; the body influences the motors and the motors influence the body. The body represents the beating part, made of microtubules and the wrapping material: in the real cell this part is subject to elasticity and hydrodynamics constraints, but for the purpose of a simplified model it can be strongly reduced to few variables characterizing the configuration of the traveling wave. The motors constitute the active source of energy for the excitation of the elastic body, they can also be modelled in more or less realistic ways. 

{\color{red}The model we address in the present work (JPK, see Sec.~\ref{JPK}) belongs to the class of $1+N$ models: the mechanical body is considered rigid and therefore is reduced to a single configurational variable $X(t)$, subject to a restoring elastic force with coefficient $k$ and viscous damping $\xi$, while the $N$ motors, located along the filament, are each described by a binary variable $s_i \in \{0,1\}$ encoding their active (bound) or inactive (unbound) state. The switching rates between the two states depend on $X(t)$, closing a 
feedback loop that sustains spontaneous oscillations. A nearest-neighbour coupling $K \geq 0$ between adjacent motors is additionally introduced, modifying the transition rates and favouring configurations in which neighbouring motors share the same state. This is a refinement of the J\"ulicher-Prost model (JP~\ref{JP}), which falls in the same 
category and which is recovered when $K=0$. The reduction of the multifilament axoneme to a single filament and its elastic energy landscape to a rigid potential modulated by a single effective degree of freedom,  originally introduced in the Julicher-Prost papers, is justified by experimental evidence - eg. from principal component analysis showing that the flagellar dynamics is low-dimensional (for instance in~\cite{Ma2014} they show that the $95\%$ of movement of the flagellar centerline  can be captured by the first two shape modes of its tangent angle). Already the observation of a stable limit cycle suggests a simple description in terms of a minimal oscillator model.}

Note that in  Appendix~\ref{hom} we also revisit a a $1+1$ model, i.e. both body and motors are represented by a single variable. We decided to summarise with some details this simpler model because it is pedagogical for the more detailed models~\cite{howard2009mechanical}.

We anticipate that, from a qualitative point of view (e.g. looking at the shape of the limit cycle), the effect of increasing the coupling $K$ among neighbouring motors is not very different from reducing the relative elasticity of the body, i.e. the parameter $k$ or its rescaled adimensional counterpart $\nu=k/(\xi\omega)$ in the original, uncoupled, JP model. As already observed in~\cite{guerin2011dynamical}, when $k$ is large the limit cycle is smooth while reducing $k$ leads to larger excursions and - consequently - a bi-stable cycle shape. The same happens in the $1+1$ ("homogeneous") model, discussed in the Appendix.  We recall the difference between $k$ and $K$: the elasticity $k$ represents the action of an external harmonic potential confining the motion of the filament, while the coupling parameter $K$ represents the tendency to be in the same state for adjacent motors. 

\subsection{The J\"ulicher-Prost model with a spatial profile}
\label{JP}

\begin{figure}[h!!]
    \centering
    \includegraphics[width=\linewidth]{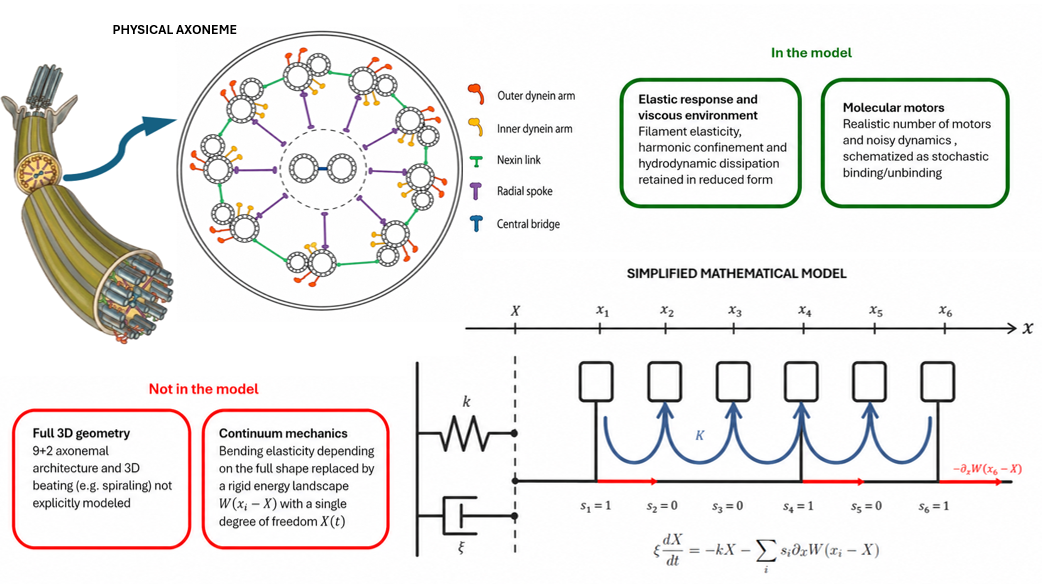}
    \caption{{\color{red}Schematic representation of the axonemal structure and its corresponding simplified mathematical model used in this study, highlighting the aspects captured by the model and those it neglects. In the mathematical description, molecular motors interact with the filament only when attached and exert forces derived from the potential $W(x-X)$. } }
    \label{fig:coupled}
\end{figure}

With respect to a homogeneous model where the state of motors is described by the number of active ones only (see for instance the model described in  Appendix~\ref{hom}), introducing $N$ variables for the motor state allows for more realism and, most importantly, for the possibility of observing fluctuations. 
A model with emerging collective dynamics of $N$ motors coupled to a filament, represented by a collective coordinate $X(t)$, has been introduced in~\cite{Julicher1997}. It was built upon a previous work~\cite{julicher1995cooperative}, where the same basic ingredients were already present but - in the absence of a confining elastic external force - the model illustrated indefinite directed motion instead of oscillations. The original model therefore displayed a transition from rest to motion, in the shape of a discontinuous transition,  at increasing ATP concentration (a value that in the model represents the breaking of detailed balance). In~\cite{Julicher1997} the presence of the elastic confinement converted the direct motion into oscillations, leading to a Hopf-like bifurcation from a fixed point to limit cycle. The model was then studied in details in~\cite{guerin2011dynamical} where it was shown generically that the dynamical instabilities are associated to a non-monotonic force velocity relation. Bidirectional motion in this model (due to finite size fluctuations and without elastic confinement) has been studied in~\cite{badoual2002bidirectional,guerin2011motion,guerin2011bidirectional}.

In the JP model, see Fig.~\ref{fig:coupled}, the system consists of an ensemble of $N$ independent motors that cyclically bind to a common filament (the ``backbone’’), which is considered rigid, whose collective position is described by $X(t)$.  Each motor $i$ is at position $z_i$  and can be in a bound state ($s_i=1$) or an unbound state ($s_i=0$), interacting with the filament through a potential which is periodic with period $\ell$:
\begin{equation} \label{eq:potential}
    W(x) = U_0 \left[1 - \cos\!\left(\frac{2\pi x}{\ell}\right)\right],
\end{equation}
where $x$ is the relative position between the motor position and the filament coordinate, see below.
Motors can be thought as regularly spaced along the filament but with a period which is incommensurate with $\ell$: the precise form of $z_i$ is however not important, the only necessary ingredient is that in the limit of $N\to\infty$ they have a homogeneous density ($\ell$ does not scale with $N$).
The binding and unbinding rates satisfy the uniform-rate condition $\omega_{\text{on}}+\omega_{\text{off}}=\Omega$, which simplifies the analytical treatment while preserving the essential physical features.  
They take the following analytical form:
\begin{equation}
    \omega_{\text{on}}(x) = \Omega\!\left[\eta - \alpha \cos\!\left(\frac{2\pi x}{\ell}\right)\right],
    \qquad
    \omega_{\text{off}}(x) = \Omega - \omega_{\text{on}}(x).
    \label{eq:rates_uncoupled}
\end{equation}
The parameter $\eta$ is the fraction of bound motors averaged over $x$ and is sometimes called the duty ratio:
\begin{equation}
    \eta=\frac{1}{\ell}\int_0^\ell dx \frac{\omega_{on}(x)}{\omega_{on}(x)+\omega_{off}(x)} = \frac{1}{\ell \Omega}\int_0^\ell dx\, \omega_{on}(x).
\end{equation}

The motion of the filament is governed by viscous and elastic forces, characterized respectively by the drag coefficient $\xi$ and the elastic coefficient $k$, as well as by active forces derived from the potential \eqref{eq:potential}.  
Introducing the dimensionless parameters
\[
    \nu = \frac{k}{\xi \Omega}, \quad
    \gamma = \frac{2\pi^2 \alpha U_0 N}{\Omega \ell^2 \xi},
\] and considering $\ell$ and $1/\Omega$ as units of space and time, 
the equation of motion becomes:
\begin{equation}
    \frac{dX}{dt} = -\nu X + \frac{\gamma}{\pi \alpha N}
    \sum_i s_i\sin\!\left[2\pi (z_i - X)\right].
    \label{eq:Xdot_dimless}
\end{equation}
We highlight, in passing, that the force exerted by the motors is evaluated at the position $x_i=z_i-X(t)$ i.e. that it takes into account the time-dependent shift $X(t)$ of the filament. For the same reason the transition rates $\omega_{on}$ and $\omega_{off}$ must be evaluated at the positions $x_i=z_i-X(t)$ (explicit in~\cite{guerin2011bidirectional,Ma2014}, but also taken into account in~\cite{julicher1995cooperative,Julicher1997,guerin2011dynamical}).
We also underline that in the following we study the model when $N$ changes but $\gamma$ is kept fixed (analogous to a rescaling $U_0 \sim 1/N$), similarly to what done in~\cite{Ma2014}. This implies that apparently the force is of order $1$ and $N$ has an effect on its fluctuations only. We say {\em apparently} because the real impact of the force depends upon the state of coordination of the motors, which is influenced by $N$ and the coupling $K$ (introduced below).

In Section~\ref{analysis} we describe - directly applied to our model - an analytical strategy for large $N$ which was already proposed by the authors~\cite{julicher1995cooperative,Julicher1997} and in successive works~\cite{guerin2011dynamical,Ma2014} for the original JP model which is a particular case ($K=0$). The main results of the study of the model at $K=0$ are briefly recalled here. The analytical strategy is based upon expanding in spatial Fourier modes the density of motors, see Eq.~\eqref{expansion}. Then one notes that the two modes at the largest wavelength, characterized (in the  frame of reference where the motor profile is comoving with $X(t)$) by coefficients $\tilde a_1(t),\tilde b_1(t)$, dominate the description of the filament. These modes obey the following  stochastic differential equations, coupled with $X(t)$:
\begin{align} \label{sde0}
    \dot{\tilde a}_1 &= -(\tilde a_1 + 1 - \gamma \tilde b_1^2 + 2\pi\nu \tilde b_1 X) + \zeta_a(t), \\
    \dot{\tilde b}_1 &= -(\tilde b_1 + \gamma \tilde b_1 \tilde a_1 - 2\pi\nu \tilde a_1 X) + \zeta_b(t), \\
    \dot{X} &= \frac{\gamma}{2\pi} \tilde b_1 - \nu X.
\end{align}
The noises~\cite{Ma2014} $\zeta_a,\zeta_b$ are Gaussian, with
\begin{align}
 \langle \zeta_i(t)\zeta_j(t') \rangle &= 2 D_i \delta_{ij}\delta(t - t'),\\
    D_a &= \frac{1}{2N}\left[\frac{4\eta(1-\eta)}{\alpha^2} - 3\right], \\
    D_b &= \frac{1}{2N}\left[\frac{4\eta(1-\eta)}{\alpha^2} - 1\right].
\end{align}

Neglecting the noises, one has a dynamical system that has a unique fixed point at $\tilde a_1^*=-1$, $\tilde b_1^*=0$ and $\dot X=0$ with $X^*=0$. The linearised equations for small force $f=\gamma \tilde b_1/(2 \pi)$ and small $X$ read
\begin{align}\label{linJP}
    \dot f=-f+\gamma \dot X,\\
    \dot X=-\nu X+f.
\end{align}
These equations can be cast into the following second-order equation for X:
\begin{equation}
    \ddot X-\epsilon \dot X+\nu X=0,
\end{equation}
where we have defined the parameter $\epsilon=\gamma-1-\nu$. When $\epsilon<0$ the fixed point is stable, when $\epsilon>0$ a limit cycle is reached. The second-order equation allows us also to estimate the oscillation frequency at the bifurcation point $\omega_0=\omega_c=\sqrt{\nu}$. Therefore the natural oscillation frequency of this flagellum model increases with the rescaled elastic coefficient $\nu$.

In the Appendix of~\cite{Ma2014} the stochastic differential equations, Eqs~\eqref{sde0} have been reduced to a stochastic Hopf normal form close to the bifurcation.  
In this regime, the oscillation phase undergoes stochastic diffusion characterized by a coefficient $D$, which quantifies the temporal loss of coherence, while the quality factor $Q$ measures the regularity of the motion: large $Q$ values correspond to more stable and coherent oscillations.  
These quantities take the form:
\begin{equation}
    D = \left[1 + \left(\frac{\omega_1}{\mu}\right)^2\right] D_b,
    \qquad
    Q = \frac{\omega_0}{2D}.
    \label{eq:Q}
\end{equation}
with explicit parameters:
\[
\mu = \frac{3\pi^2 \Omega \nu (1 + 2\nu)}{2(1 + 4\nu)}, \quad \Lambda = \frac{1}{\pi^2} \cdot \frac{\epsilon(1 + 4\nu)}{3\nu(1 + 2\nu)},
\]
\[
\omega_c = \Omega \sqrt{\nu}, \quad \omega_1 = -\mu \sqrt{\nu}/(1 + 2\nu),  \quad \omega_0=\omega_c-\omega_1\Lambda.
\]
We also recall that $\sqrt{\Lambda}$ represents the amplitude of oscillations and $\omega_0$ their frequency.
It follows that $Q \propto N$, confirming that oscillation coherence increases linearly with the number of  motors~\cite{Sharma}.  
It is also useful to note that $Q \propto 1/D_b$, a relation that will be recalled later when discussing the effects of coupling.

\subsection{The model with nearest-neighbour interactions}
\label{JPK}

In \cite{Maggi2022}, Maggi and collaborators proposed an extension of the JP model introducing local interactions between adjacent motors, modeled by a quadratic potential:
\begin{equation}
    U_{\text{bind},i} = K\left[(s_i - s_{i+1})^2 + (s_i - s_{i-1})^2\right],
\end{equation}
where $K$ quantifies the coupling strength.  
Minimising this energy term favors configurations in which neighboring motors share the same state, promoting local coordination. 

The coupling energy modifies the transition rates. How the rates depend upon this energy in the original paper and how differently is considered in this paper is explained in the Appendix, see~\ref{rates}. Here we give only the expressions used in the present work.
Assuming that the effect of the potential appears as an exponential correction $\omega_{\text{on},i}^{(0)} \, e^{-\Delta U}$, one obtains:
\begin{align}
    \omega_{\text{on},i} &= 
    \Omega\!\left[\eta - \alpha \cos\!\left(\frac{2\pi (z_i - X)}{\ell}\right)\right]
    e^{-2K(1 - s_{i-1} - s_{i+1})}, \\
    \omega_{\text{off},i} &=
    \Omega - \Omega\!\left[\eta - \alpha \cos\!\left(\frac{2\pi (z_i - X)}{\ell}\right)\right]
    e^{-2K(1 - s_{i-1} - s_{i+1})}.
\end{align}
This formulation preserves the uniform-rate condition and introduces cooperative dynamics: a motor tends to bind if its neighbors are bound, and to unbind otherwise.

\section{Functional Fokker-Planck equation} \label{ffp}

As a starting point, we derive the functional Fokker–Planck equation in the model with nearest-neighbor couplings in the large $N$ limit.
To achieve this goal, we adopt the following strategy, similar but not identical to the derivations found in~\cite{Julicher1997,guerin2011dynamical,Ma2014}. The filament is divided into $m$ equally spaced sites, each containing $N/m$ motors. We introduce the probability $P(\{n_i\}, X, t)$ of finding the vector of motor states ($i=1 ... N$)  in configuration $\{n_i\}$, while the filament is at position $X$ at time $t$.  
In the limit $N \gg 1$, one obtains a general Fokker–Planck equation valid for arbitrary choices of the rates $\omega_{\text{on/off}}$.
\begin{align} \label{eq: FP Julicher}
    \frac{\partial P}{\partial t}
    =& -\frac{\partial}{\partial X}
    \left[
        \left(
            \sum_i \frac{W'(z_i - X) n_i}{\xi N}
            - \nu X
        \right) P
    \right]+ \nonumber\\
    &+ \sum_i \frac{\partial}{\partial n_i}
    \Big[
        \big(
            \omega_{\text{off}}(z_i - X) n_i 
            - \omega_{\text{on}}(z_i - X)(N/m - n_i)
        \big) P
    \Big]+ \nonumber\\
    &+\frac{1}{2}
    \sum_i \frac{\partial^2}{\partial n_i^2}
    \Big[
        \big(
            \omega_{\text{off}}(z_i - X) n_i 
            + \omega_{\text{on}}(z_i - X)(N/m - n_i)
        \big) P
    \Big].
\end{align}
Since the variables $s_i$ are discrete, the model is reformulated in terms of the mean density of bound motors at site $i$, defined as $\rho_i = n_i/(N/m\ell)$.  Except for edge motors (negligible in number), the neighbors $s_{i-1}$ and $s_{i+1}$ belong to the same site $j$, one obtains, in the mean-field approximation:
\[
    \langle s_i \rangle = \langle s_{i\pm1} \rangle = \rho_j \ell.
\]
The mean transition rates for site $i$ thus become:
\begin{align}
    \omega_{\text{on},i} &= 
    \Omega\!\left[\eta - \alpha \cos\!\left(\frac{2\pi (z_i - X)}{\ell}\right)\right]
    e^{-2K(1 - 2\rho_i \ell)}, \\
    \omega_{\text{off},i} &=
    \Omega - \Omega\!\left[\eta - \alpha \cos\!\left(\frac{2\pi (z_i - X)}{\ell}\right)\right]
    e^{-2K(1 - 2\rho_i \ell)}.
\end{align}
We stress the fact that, though the notation remains the same, the meaning of the rate is slightly different, passing from a variable at motor level to a site-averaged variable.
In the continuum limit $\rho_i \to \rho(z)$, one obtains:
\begin{align}
    \omega_{\text{on}}(z - X) &=
    \Omega\!\left[\eta - \alpha \cos\!\left(\frac{2\pi (z - X)}{\ell}\right)\right]
    e^{-2K(1 - 2\rho(z)\ell)}, \\
    \omega_{\text{off}}(z - X) &=
    \Omega - \omega_{\text{on}}(z - X).
\end{align}
In the continuum limit $\rho_i \to \rho(z)$ a functional Fokker-Planck replaces the discrete one. The functional formulation reads:
\begin{align}
    \frac{\partial P}{\partial t} =
    &- \frac{\partial}{\partial X}[v(X,[\rho])P]
    + \int_0^{\ell} dz \, \frac{\delta}{\delta \rho(z)}[A(z,X,\rho)P] \nonumber\\
    &+ \frac{1}{2N} \int_0^{\ell}\!\! dz \int_0^{\ell}\!\! dy \, 
    \delta(z-y)\frac{\delta^2}{\delta\rho(z)\delta\rho(y)}[B(z,X,\rho)P],
    \label{eq:FP functional}
\end{align}
where
\begin{align}
    v(X,[\rho]) &= \frac{1}{\xi}\int_0^{\ell} W'(z - X)\rho(z)\,dz - \nu X, \\
    A(z,X,\rho) &= \omega_{\text{off}}(z - X)\rho(z)
    - \omega_{\text{on}}(z - X)\!\left[\frac{1}{\ell} - \rho(z)\right], \\
    B(z,X,\rho) &= \omega_{\text{off}}(z - X)\rho(z)
    + \omega_{\text{on}}(z - X)\!\left[\frac{1}{\ell} - \rho(z)\right].
\end{align}


\section{Linear analysis for small coupling} \label{analysis}

The local density of bound motors can be expanded as a Fourier series:
\begin{equation}
    \rho(z,t) = \alpha a_0(t) \label{expansion}
    + \alpha\!\sum_{n=1}^{\infty}\left[a_n(t)\cos\!\left(n \frac{2\pi z}{\ell}\right)
    + b_n(t)\sin\!\left(n \frac{2\pi z}{\ell}\right)\right],
\end{equation}
where $a_n$ and $b_n$ are the Fourier components and $\alpha$ is the same parameter introduced in the rates definition. In the following we consider the deterministic (noise-free) limit, neglecting the noise terms in the functional Fokker–Planck equation.

At first order in $K$, the system of equations for the evolution of the filament position and the Fourier modes becomes:
\begin{center}
\begin{align} \label{eq:sistema}
     \dot X &= \frac{\gamma}{2 \pi} \ell^2\Omega(b_1c_X - a_1s_X) - \nu X, \\
 \nonumber    \dot a_0 &= -\left[ \Omega(1-4\eta K)a_0 - \frac{\Omega}{\ell}(1-2K)\frac{\eta}{\alpha} +2\Omega K \alpha(c_Xa_1+s_Xb_1)\right],\\
 \nonumber    \dot a_1 &= -\left[\Omega(1-4\eta K)a_1+\frac{\Omega}{\ell}(1-2K)c_X+2\Omega K \alpha(2c_Xa_0+c_Xa_2+s_Xb_2)\right], \\
 \nonumber    \dot b_1 &= -\left[\Omega(1-4\eta K)b_1+\frac{\Omega}{\ell}(1-2K)s_X+2\Omega K \alpha(2s_Xa_0+s_Xa_2+c_Xb_2)\right],\\
 &\nonumber \text{for } n \geq 2,\\
  \nonumber   \dot a_n &= -\left[\Omega(1-4\eta K)a_n+2\Omega K \alpha\left((a_{n+1}+a_{n-1})c_X+((b_{n+1}-b_{n-1})s_X)\right)\right],\\
  \nonumber   \dot b_n &= -\left[\Omega(1-4\eta K)b_n+2\Omega K \alpha\left((a_{n+1}+a_{n-1})s_X+((b_{n+1}+b_{n-1})c_X)\right)\right],
\end{align}
\end{center}
with the definitions $c_X = \cos(2\pi X/\ell)$ and $s_X = \sin(2\pi X/\ell)$. The resulting dynamics is physically equivalent to that of JP, although the equations are not written in the same form, since they are derived in a different reference frame. In \eqref{comoving} below, we rewrite the dynamics in the frame co-moving with the flagellum; in that representation, setting K=0 recovers exactly the JP equations (i.e. Eqs.~\eqref{sde0}, neglecting noise).

To simplify the expressions we introduce dimensionless variables by setting $\ell = 1$ and $\Omega = 1$, so that all quantities are expressed in natural units.  
We then expand the Fourier modes to first order:
\[
a_n = a_n^{(0)} + K a_n^{(1)}, \qquad b_n = b_n^{(0)} + K b_n^{(1)}.
\]

At zeroth order we recover the original JP model:
\begin{center}
\begin{align}
\dot a_0^{(0)} &= -\Big[a_0^{(0)} - \frac{\eta}{\alpha}\Big],\\
\nonumber \dot a_1^{(0)} &= -\Big[a_1^{(0)} + c_X\Big],\\
\nonumber \dot b_1^{(0)} &= -\Big[b_1^{(0)} + s_X\Big],\\
n \geq 2:\nonumber\\
\nonumber \dot a_n^{(0)} &= -a_n^{(0)},\\
\nonumber \dot b_n^{(0)} &= -b_n^{(0)}.
\end{align}
\end{center}

At first order the system becomes:
\begin{center}
\begin{align}
\dot a_0^{(1)} &= -\Big[- 4\eta\, a_0^{(0)} + a_0^{(1)} + 2\frac{\eta}{\alpha} + 2\alpha(c_X a_1^{(0)} + s_X b_1^{(0)})\Big],\\
\nonumber \dot a_1^{(1)} &= -\Big[ - 4\eta\, a_1^{(0)} + a_1^{(1)} - 2 c_X + 2\alpha(2 c_X a_0^{(0)} + c_X a_2^{(0)} + s_X b_2^{(0)})\Big],\\
\nonumber \dot b_1^{(1)} &= -\Big[- 4\eta\, b_1^{(0)} + b_1^{(1)} - 2 s_X + 2\alpha(2 s_X a_0^{(0)} + s_X a_2^{(0)} + c_X b_2^{(0)})\Big],\\
n \geq 2:\nonumber\\
\nonumber \dot a_n^{(1)} &= -\Big[-4\eta\, a_n^{(0)} + a_n^{(1)} + 2\alpha\big((a_{n+1}^{(0)} + a_{n-1}^{(0)}) c_X + (b_{n+1}^{(0)} - b_{n-1}^{(0)}) s_X\big)\Big],\\
\nonumber \dot b_n^{(1)} &= -\Big[-4\eta\, b_n^{(0)} + b_n^{(1)} + 2\alpha\big((a_{n+1}^{(0)} + a_{n-1}^{(0)}) s_X + (b_{n+1}^{(0)} + b_{n-1}^{(0)}) c_X\big)\Big].
\end{align}
\end{center}

In the long-time regime, the variables $a_0^{(0)}$ and $a_n^{(0)}$ for $n \ge 2$ relax to their  fixed points, yielding:
\begin{center}
\begin{align}\label{eq:sistema pert}
\dot X &= \frac{\gamma}{2\pi}\big[(b_1^{(0)} + K b_1^{(1)}) c_X - (a_1^{(0)} + K a_1^{(1)}) s_X\big] - \nu X,\\
\nonumber a_0^{(0)} &= \frac{\eta}{\alpha},\\
\nonumber \dot a_1^{(0)} &= -\big[a_1^{(0)} + c_X\big],\\
\nonumber \dot b_1^{(0)} &= -\big[b_1^{(0)} + s_X\big],\\
\nonumber a_n^{(0)} &= 0,\quad b_n^{(0)} = 0 \quad \text{for } n \geq 2,\\
\nonumber \dot a_0^{(1)} &= -\Big[-4\frac{\eta^2}{\alpha} + a_0^{(1)} + 2\frac{\eta}{\alpha} + 2\alpha(c_X a_1^{(0)} + s_X b_1^{(0)})\Big],\\
\nonumber \dot a_1^{(1)} &= -\Big[-4\eta\, a_1^{(0)} + a_1^{(1)} - 2 c_X + 4\eta c_X\Big],\\
\nonumber \dot b_1^{(1)} &= -\Big[-4\eta\, b_1^{(0)} + b_1^{(1)} - 2 s_X + 4\eta s_X\Big],\\
\nonumber \dot a_2^{(1)} &= -\Big[a_2^{(1)} + 2\alpha\big(a_1^{(0)} c_X - b_1^{(0)} s_X\big)\Big],\\
\nonumber \dot b_2^{(1)} &= -\Big[b_2^{(1)} + 2\alpha\big(a_1^{(0)} s_X + b_1^{(0)} c_X\big)\Big],\\
\nonumber a_n^{(1)} &= 0,\quad b_n^{(1)} = 0 \quad \text{for } n \geq 3.
\end{align}
\end{center}

The system displays a clear hierarchical structure: first-order equations are directly driven by the zeroth-order solutions, which act as forcing terms, while the zeroth-order dynamics is completely unaffected by first-order corrections, as expected from a perturbative expansion.  
All equations contain a negative linear relaxation term of the form $\dot y = -y + \ldots$, implying dissipative dynamics. As a consequence, the system possesses dynamical attractors (fixed points or limit cycles) that determine its long-time behaviour.  
Moving from zeroth to first order in $K$ also makes the equations for $n\ge 2$ nontrivial: the truncation, which in the uncoupled limit effectively occurs at $n=1$, naturally shifts to $n=2$. This reflects the structure of the unperturbed problem, where the linear term in $K$ couples neighbouring modes, linking $a_n^{(1)}$ and $b_n^{(1)}$ to $a_{n-1}^{(0)}$ and $b_{n-1}^{(0)}$. We also highlight that the structure of the linear system is highly non-reciprocal (e.g. the coefficient of variable $a_n$ appearing in $\dot a_m$ is different when $n$ and $m$ are swapped), a fact which is in line with the active nature of the model under investigation~\cite{loos2020irreversibility,fruchart2023odd,hickey2023nonreciprocal,klapp2023non,ishimoto2023odd}. The non-reciprocal structure of the linear system is a direct consequence of the full model being non-potential.

\begin{figure}[h!]
    \centering
    \includegraphics[width=0.48\textwidth]{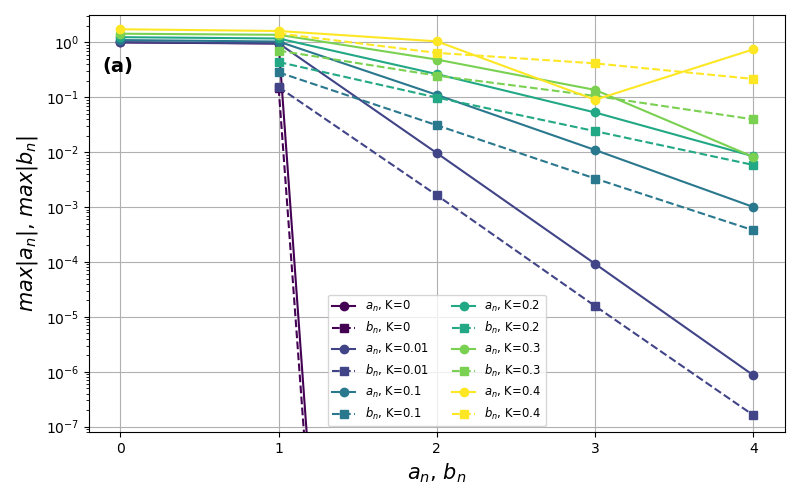}
        \includegraphics[width=0.48\textwidth]{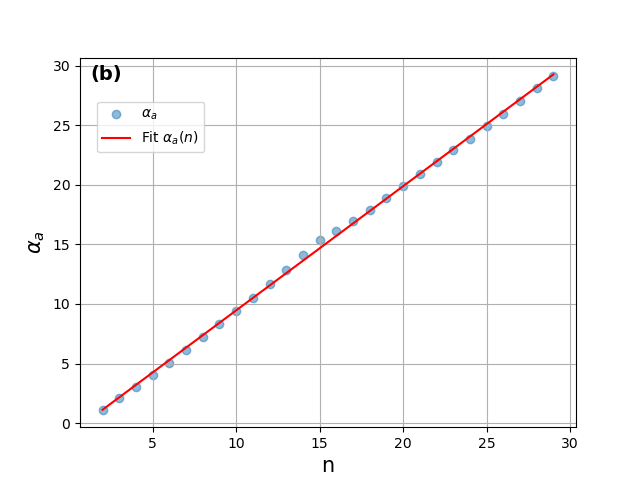}
    \hfill
        \includegraphics[width=0.99\textwidth]{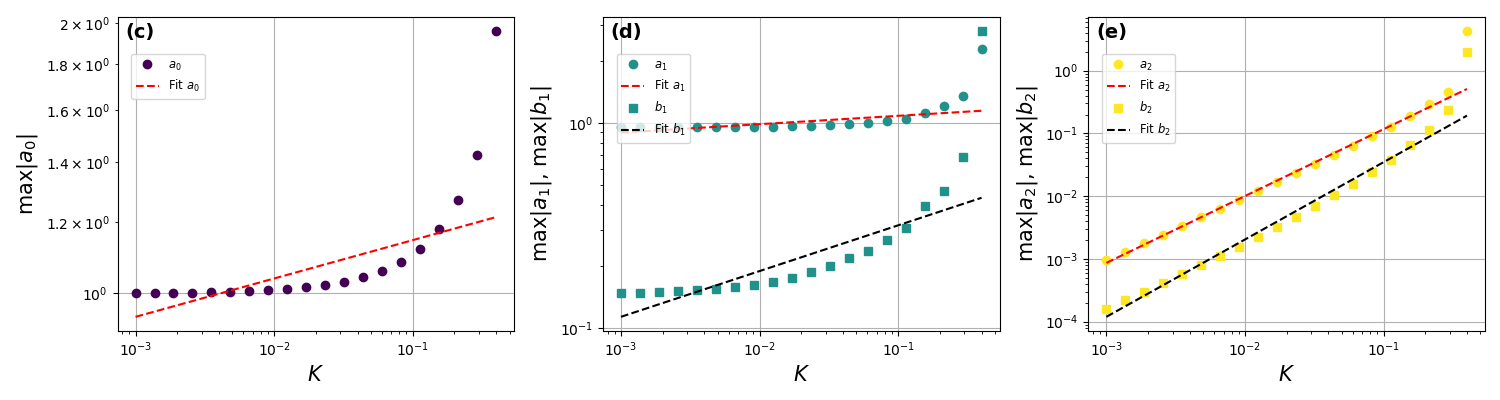}
    \caption{Numerical results about the decay of the maximum of the Fourier mode from the solution of the system \eqref{eq:sistema}.
    (a) maxima as a function of the mode index $n$ (the dashed line connecting square symbols corresponds to the coefficients $b_n$); (b) fitted values of the exponent $\alpha_a(n)$, governing the power-law scaling $\sim K^{\alpha_a(n)}$, shown as a function of $n$;
    (c-e) maxima of the first Fourier coefficients as functions of $K$; dashed lines indicate power-law fits $\sim K^{\alpha_a(n)}$ and $\sim K^{\alpha_b(n)}$.}
        
    \label{fig:Fourier}
\end{figure}

An analysis of the full system \eqref{eq:sistema} can be performed via numerical integration.
For our purposes, the integration is done by truncating the system at index $n=60$, with a time step $dt=0.001$, up to $t=300$, using the same parameters reported in Table \ref{tab:params}, shown in the next Section. As displayed in Figure~\ref{fig:Fourier}, the maxima of the oscillating coefficients $a_n(t),b_n(t)$ after a certain transient (taken as 90\% of the total time) differ by orders of magnitude depending on $K$.  Most importantly there is a clear difference between the cases $K=0$ and $K>0$: in the former case $a_n$ and $b_n$ fall to zero starting with $n=2$, while in all other cases $a_n$ and $b_n$ decrease exponentially with $n$: this is a clear evidence that even small couplings modify the structure of the density profile.  

For Fourier coefficients with larger indices, the double-logarithmic scale plot reveals a power-law dependence of the maxima on $K$, at least for $K<0.4$. At $K=0.5$, the sign of $(1-4\eta K)$ changes, causing the coefficients to increase exponentially instead of decreasing; thus, the perturbative approach is no longer physically meaningful.   For each coefficient, the exponent of the power law $\sim K^{\alpha_a(n)}$-$\sim K^{\alpha_b(n)}$ is extracted, and a linear dependence of these exponents on the coefficient index $n$ is observed and fitted (Fig \ref{fig:Fourier})(only for $n\geq 2$).  
The obtained dependencies are:
\begin{equation}
    \alpha_a(n) = 1.04\, n - 0.97, \qquad 
    \alpha_b(n) = 1.04\, n - 0.86.
\end{equation}
{\color{red}Additional Fourier coefficients with larger indices are reported in the Appendix~\ref{decay}, where they further confirm the persistence of the exponential decay regime for coefficients with $n \geq 2$.}

\subsection{Comoving frame of reference}


To maintain consistency with JP framework, in particular with the relation between noise and quality factor, we study the system also in the frame co-moving with the filament by defining
$\tilde{\rho}(x) = \rho(x + X(t))$ and consequently the coefficients $\tilde a_n$ and $\tilde b_n$ for the corresponding Fourier decomposition. 
Inserting this transformation into the Fokker–Planck equation yields:
\begin{center}
\begin{align} \label{comoving}
     \dot X &= \frac{\gamma}{2 \pi} \tilde b_1 - \nu X ,\\
     \nonumber\dot {\tilde a}_0 &= -\left[ \Omega(1-4\eta K)\tilde a_0 - \frac{\Omega}{\ell}(1-2K)\frac{\eta}{\alpha} +2\Omega K \alpha \tilde a_1\right],\\
     \nonumber\dot {\tilde a}_1 &= -\left[\Omega(1-4\eta K)\tilde a_1+\frac{\Omega}{\ell}(1-2K)+2\Omega K \alpha(2\tilde a_0+\tilde a_2)-\frac{\gamma \tilde b_1^2}{\ell}+\frac{2\pi \nu \tilde b_1}{\ell}X\right], \\
     \nonumber\dot {\tilde b}_1 &= -\left[\Omega(1-4\eta K)\tilde b_1+2\Omega K\alpha \tilde b_2  +\frac{\gamma \tilde b_1 \tilde a_1}{\ell}-\frac{2\pi \nu \tilde a_1}{\ell}X\right],\\
     \nonumber \dot {\tilde a}_n &= -\left[\Omega(1-4\eta K)\tilde a_n+2\Omega K \alpha(\tilde a_{n+1}+\tilde a_{n-1}) -\frac{\gamma n\tilde b_1\tilde b_n}{\ell}+\frac{2\pi \nu n\tilde b_n}{\ell}X\right],\\
     \nonumber\dot {\tilde b}_n &= -\left[\Omega(1-4\eta K)\tilde b_n+2\Omega K \alpha(\tilde b_{n+1}+\tilde b_{n-1})+\frac{\gamma n \tilde b_1\tilde a_n}{\ell}-\frac{2\pi \nu n \tilde a_n}{\ell}X\right],
\end{align}
\end{center}
which reduces to the uncoupled case when $K=0$. An important result, already mentioned before but more evident in this frame of reference, is the fact that, for the fixed point of the system, when $K=0$ one has $\tilde a_n=0$ for $n \ge 2$, while when $K>0$ such coefficients have non-zero values (but smaller and smaller with growing $n$).

A fundamental result, that is more clear in the comoving frame of reference, is that the rate of self-decay (the coefficient of $-a_n$ in $\dot a_n$ or the coefficient of $-b_n$ in $\dot b_n$, in each mode, {\em decreases} with $K$ as $\Omega(1-4 \eta K)$. The correction $(1-4 \eta K)$, at all orders reveals a main effect of $K$ for the dynamics, which becomes slower and slower with larger values of the coupling. This observation is consistent with the estimate of the oscillation frequency (see next subsection) that decreases as $\approx \sqrt{\nu (1-4 \eta K)}$ and the measurement of the limit cycle period in simulations which increases exponentially with $K$. From the observation that for large $K$ the period grows exponentially with $K$, see Fig.~\ref{cycles}(f), we conjecture  - at least phenomenologically - that the correction $\Omega(1-4\eta K)$ can be prolonged for large $K$ as $e^{-4\eta K}$.

Reintroducing noise in the flagellar frame, using the same definitions and the same procedures as in the original model~\cite{Ma2014}, the noise intensities become:
\begin{align}
    D_{\tilde a} &= \frac{1}{2N}\left[\frac{4\eta(1 - \eta)}{\alpha^2} - 3\right] + \frac{K}{N\alpha^2} \left[10\alpha^2 + 2\alpha + 12 - 32\eta \alpha^2 - 4\eta \alpha - 16\eta^3-4\eta \right]
    =\frac{1}{N}\left(d_{\tilde a}^{(0)} + K d_{\tilde a}^{(1)}\right), \\
    D_{\tilde b} &= \frac{1}{2N}\left[\frac{4\eta(1 - \eta)}{\alpha^2} - 1\right] + \frac{K}{N\alpha^2} \left[2\alpha^2 + 2\alpha + 12\eta^2 - 8\eta \alpha^2 - 4\eta \alpha - 16\eta^3 -4\eta\right]
    = \frac{1}{N}\left(d_{\tilde b}^{(0)} + K d_{\tilde b}^{(1)}\right),
\end{align}
where $d^{(1)}$ denotes the linear correction due to coupling and $d^{(0)}$ corresponds to the same results found in~\cite{Julicher1997}.

Since the quality factor scales as $Q \propto 1/D_{\tilde b}$, the corrected expression becomes:
\begin{equation}
        Q \propto \frac{1}{D_{\tilde b}}= \frac{N}{d_{\tilde b}^{(0)} + K d_{\tilde b}^{(1)}},
\end{equation}
showing that even weak coupling directly affects the temporal coherence of the oscillations.
From this formula, in the small-coupling limit, we also expect $Q$ to receive a correction that scales linearly with $N$. 

Following the choices commonly adopted in previous works and in the simulations of this study, we set the parameters $\alpha = \eta = \tfrac{1}{2}$. With this choice, the correction $d_{\tilde b}^{(1)}$ is equal to $-6$. Consequently, in the small-coupling limit, we expect that the quality increase with increasing coupling strength $K$. Numerical result, see below, reveal that the quality in fact increases for small $K$, but for large $K$ decreases, contradicting the  prediction at linear order.

\subsection{Stability of an effective coarse-grained system}
\label{reduced}
\begin{figure}
    \centering
    \includegraphics[width=0.5\linewidth]{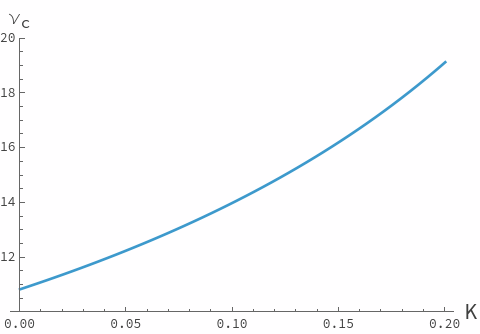}
    \caption{Critical $\nu$ as a function of $K$ from the stability analysis of the system reduced to only three variables, Eqs.~\eqref{only3}. When $\nu>\nu_c$ the fixed point is stable, while if $\nu<\nu_c$ a limit cycle appears.}
    \label{fignuc}
\end{figure}

In this subsection we discuss some approximation which is justified by the fair comparison observed with numerical simulations. The advantage is the reduction to a minimal set of degrees of freedom and the possibility of a simplified linear stability analysis.
Despite the infinite hierarchy of modes, simulations for small $K$ show that rapidly $\tilde a_0(t) \to \tilde a_0^*=\frac{1-2 K}{1-4 \eta K}\frac{\eta}{\alpha}$ and - even more rapidly - the system reaches $\tilde a_n \approx \tilde a_n^*$ and $\tilde b_n \approx 0$ for $n\ge 2$, with $\tilde a_{n+1}^* \ll \tilde a_n^*$, enabling a substantial empirical reduction.  It can be verified, numerically, that the dynamics of the reduced set $(X,\tilde a_1, \tilde b_1)$ is quite similar to the behaviour observed in the microscopic model.   This {\em empirically} reduced system reads:
\begin{align} \label{only3}
\dot X &= \frac{\gamma}{2\pi}{\tilde b}_1-\nu X\\
\nonumber \dot {\tilde a}_1&= -[q{\tilde a}_1+p+4 \alpha K {\tilde a}_0^*-\gamma \tilde b_1^2+2\pi \nu \tilde b_1 X]\\
\nonumber \dot {\tilde b}_1 &= -[q \tilde b_1 + \gamma \tilde a_1 \tilde b_1- 2 \nu \pi \tilde a_1 X].
\end{align}
with $q=1-4 \eta K$ and $p=1-2K$. 
The above system has a fixed point in $X=X^*=0$, $\tilde b_1=\tilde b_1^*=0$, $\tilde a_1=\tilde a_1^*=-(p+4K\alpha \tilde a_0^*)/q$. The equations linearised around the fixed point show that the fluctuations $\delta \tilde a_1=\tilde a_1-\tilde a_1^*$ are decoupled from the other two variables (as it happens for $K=0$), leaving 
\begin{align} \label{linred}
    \delta \dot {\tilde a}_1 &= -q\delta \tilde a_1, \\
    \dot X &=f - \nu X,\\
    \dot f &=-q f-\gamma a_1^* \dot X,
\end{align}
where $f=\gamma \tilde b_1/2\pi$. The last two equations can be recast into 
\begin{equation} \label{ddot}
    \ddot X - (\epsilon+\delta \epsilon)\dot X+ \nu (1-4 \eta K) X=0,
\end{equation}
where $\epsilon=\gamma-1-\nu$ was the parameter defined in Section~\ref{JP} and 
\begin{equation}
    \delta \epsilon= -\gamma(1+\tilde a_1^*)+4 \eta K  \approx  -2[\gamma(1-4\eta)-2\eta]K+O(K^2).
\end{equation} 
The system is unstable when $\epsilon + \delta \epsilon>0$, or equivalently $\epsilon>-\delta \epsilon$. From the point of view of physical parameters, when $K$ increases the unstable space enlarges, for instance the critical $\nu_c(K)$ (above which the fixed point is stable) increases with $K$, see Fig.~\ref{fignuc}.  Comparison of the numerical solution of the reduced system with a system reduced to a larger set of variables shows that the critical point is not exactly estimated, however its qualitative behavior is reproduced. 
Eq.~\eqref{ddot} gives also a first estimate (at small $K$) of the frequency of the limit cycle close to the transition, that is $\approx \sqrt{\nu (1-4 \eta K)}$. 

In conclusion, the effect of coupling $K>0$ (at least when it is small) is an increase of the possibility to observe limit cycles, together with a reduction of the limit cycle frequency at the critical point. This conclusion is qualitatively coherent with the observations of the numerical simulation of the microscopic model, see next Section and in particular Fig.~\ref{fig:phase trans}.



\section{Beyond the linear analysis: the effect of strong couplings} \label{numerics}

In order to investigate the behavior of the system beyond the regime of small $K$ we need to resort to numerical simulations of the microscopic stochastic model. Note that in the  expansion discussed above, the coefficients change sign when $K$ is large and $1-4\eta K<0$, which represents the limit of validity of the expansion itself. 
The results in this Section are obtained by numerically integrating the following dimensionless equations (which are the same as those presented in Section~\ref{JP}, we repeat them for the benefit of the reader):
\begin{align}
    \frac{dX}{dt} &= -\nu X +  F(t), \\
F(t) &= \sum_{i=1}^N s_i \frac{\gamma}{\pi\alpha N}
    \sin\big[2\pi(x_i - X(t))\big],
    \end{align}
where $s_i\in\{0,1\}$ is the state of motor $i$ and $x_i=i/N$ its position.  
At each time step, the state of a motor is updated with probability $p$ taking the following expressions:
\begin{equation*}
  s_i : 0\to 1 \text{ with }p=\omega_{\mathrm{on}}(x_i-X)\,dt,\qquad 
  s_i : 1\to 0 \text{ with }p=\omega_{\mathrm{off}}(x_i-X)\,dt.  
\end{equation*}
It is useful to note that, in the coupled case, the typical time of a transition scales as $1/\omega_{\mathrm{on}}$, which scales as $\sim e^{- c K}$ with $c$ of order $\sim 1$, depending on the state of neighboring motors.  
For large values of $K$, this implies that transition times become extremely short, much shorter than the integration time step $dt$ accessible in simulations.  
For this reason, the range of simulated values of $K$ is restricted to $K \leq 3$.  
Within this range, we observe that the qualitative behavior of the model does not depend on the specific update scheme used for the motor states, whereas deviations appear for larger values of $K$.
{\color{red}The integration of Eq.~\ref{eq:Xdot_dimless} is performed using a fourth–order Runge–Kutta scheme, and the parameters employed are listed in Tab.~\ref{tab:params}. The binding–unbinding process is treated using a fixed time–step renewal scheme (with the same $\Delta t$ as for the continuous part of the dynamics). At each step, transition probabilities are evaluated, and switching events are implemented via pseudo–random sampling. A Gillespie algorithm is not used, as the continuous dynamics already requires a fixed small time step.}

\begin{table}[h!]
\centering
\begin{tabular}{c c c}
\hline
Parameter & Description & Value \\
\hline
$\gamma$ & Active force & $1.2\pi^2$ \\
$\nu$    & Elastic constant & $10$ \\
$\eta$   & Mean fraction of active motors (for $K=0$) & $0.5$ \\
$\alpha$ & Transition–rate parameter & $0.5$ \\
$\Delta t$ & Time step & $0.001$ \\
\hline
\end{tabular}
\caption{Simulation parameters.}
\label{tab:params}
\end{table}

\begin{figure}[h!]
        \includegraphics[width=0.48\linewidth]{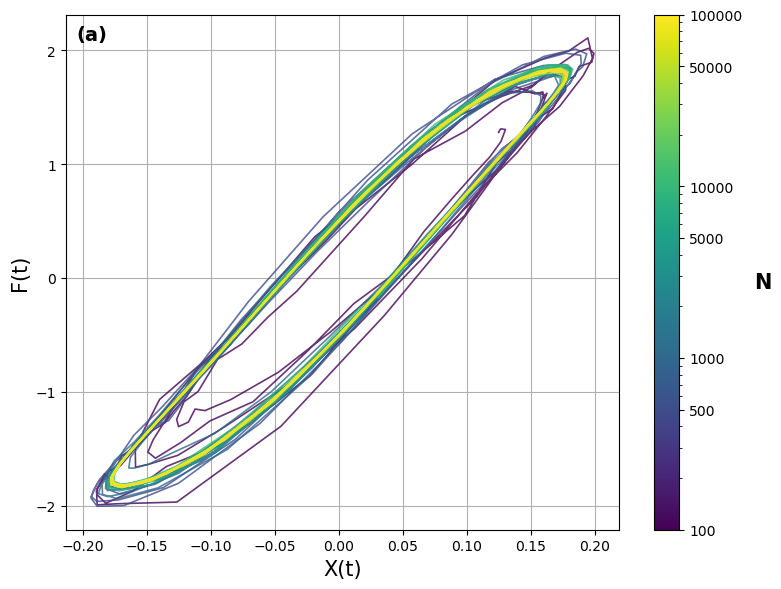}
        \includegraphics[width=0.48\linewidth]{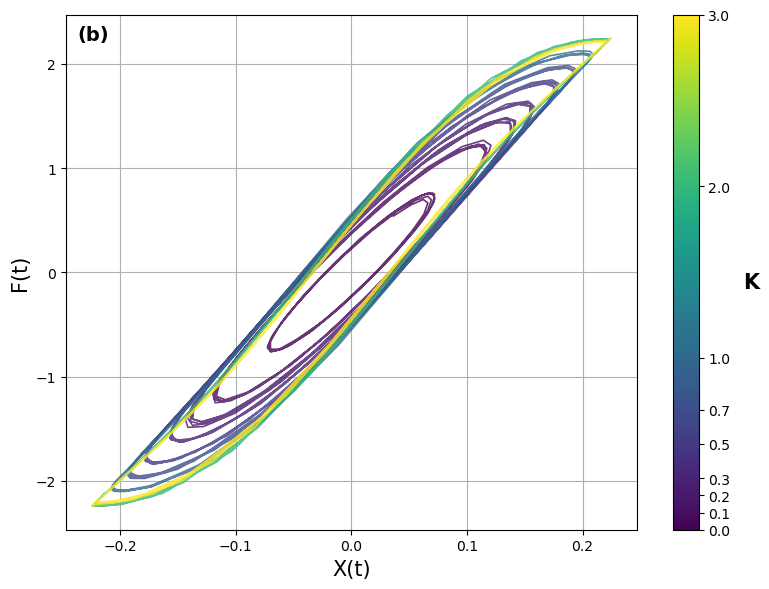}
         \includegraphics[width=0.48\linewidth]{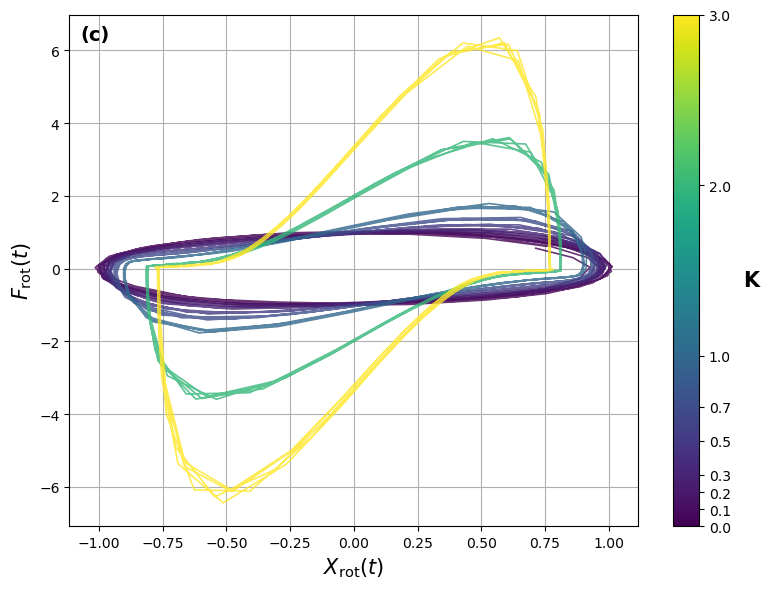}
        \includegraphics[width=0.47\linewidth]{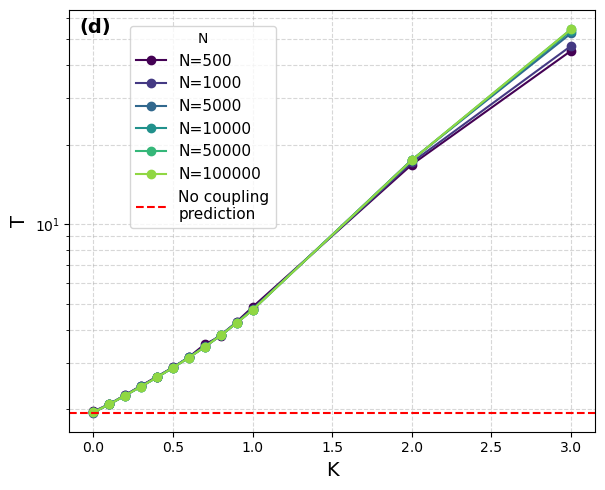}
    \caption{
    (a,b): Force-position limit cycle in numerical simulations with increasing $N$ at fixed $K=0.5$ (a) and effect of increasing coupling $K$ at fixed $N=5\cdot10^4$ (b).
    (c) PCA rotation of $(X,F)$ variables: effect of varying $K$ at fixed $N$.
    (d): Limit-cycle period in function of $K$ for several values of $N$.    \label{cycles}    }
\end{figure}

With the chosen parameters, a limit cycle is observed for all values of $N$ we have studied; for instance, it is well visible in the $(X,F)$ plane. See Fig.~\ref{cycles} a and b, representing spontaneous irreversible oscillations of the filament configuration $\{X,\sigma_1,...,\sigma_N\}$, fully consistent with the theoretical predictions. We first note that increasing the number of motors $N$ has the qualitative effect of suppressing the noise and sharpening the cycle.
More interesting is the fact that the shape of the cycle substantially changes when $K$ is increased.
For $K=0$, the cycle is nearly elliptical; when interactions between motors are introduced through a positive $K$, the cycle becomes distorted with larger displacements of $X(t)$ and forces $F(t)$. We explain such larger displacements and forces as a consequence of the slower dynamics, according to the previous estimate of the decay rate of the modes, decreasing with $K$: a slower dynamics of the density profile induces more persistent effects of the motors and therefore a more effective driving force. This is reflected also in the the measurement of the period of the limit cycle in numerical simulations, see Fig.~\ref{cycles}f: we observe an exponential increase of the period $2 \pi/\omega_0$ (where $\omega_0$ is the angular frequency of flagellar beating, see below) as a function of $K$, coherent with the predicted decrease of the frequency at small $K$, estimated in Section~\ref{reduced}. 

Since the limit cycle is neither centered nor aligned with the coordinate axes, the angular variable $\theta(t)=\arctan(F/X)$ is not well suited to track the dynamics. Instead, we introduce a more appropriate phase variable by applying a rotation by principal component analysis (PCA) that diagonalizes the covariance matrix of $(X,F)$~\cite{Ma2014}. The PCA yields two eigenvectors, $\vec{v}_0$ and $\vec{v}_1$, associated with the directions of maximal and minimal variance. The first eigenvector,
$\vec{v}_0$,
identifies the direction along which the limit cycle stretches the most. The angle $\beta$ between the original $X$ axis and $\vec{v}_0$ is therefore the rotation needed to realign the cycle along its principal directions and remove the cross–correlation between $X$ and $F$.
The rotation by $\beta$ defines two new components $(X_{\mathrm{rot}},F_{\mathrm{rot}})$, from which a well–defined phase can be constructed:
\begin{equation*}
    \cos\theta = \frac{\cos\beta\,X - \sin\beta\,F}{\sqrt{2 w_1}},\qquad
    \sin\theta = \frac{\sin\beta\,X + \cos\beta\,F}{\sqrt{2 w_0}},
\end{equation*}
where $w_0\ge w_1$ are the eigenvalues of the covariance matrix. The PCA rotation is particularly effective in the case $K=0$, see Fig.~\ref{cycles} frame c. On the contrary when $K$ increases, even after PCA rotation, the limit cycle results non-symmetric with respect to the principal axes.

\begin{figure}[h!]
    \centering
        \includegraphics[width=0.43\linewidth]{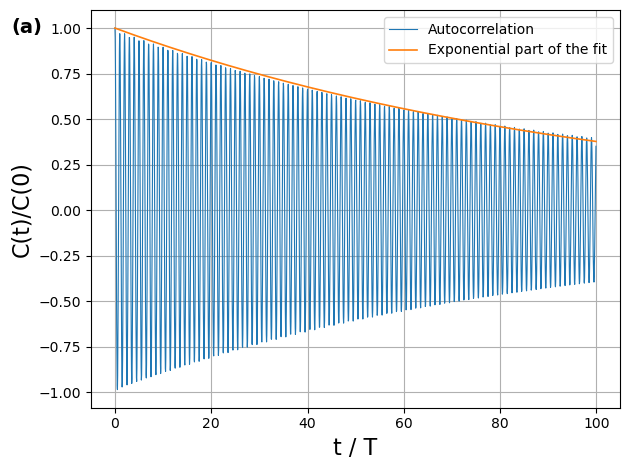}
        \includegraphics[width=0.43\linewidth]{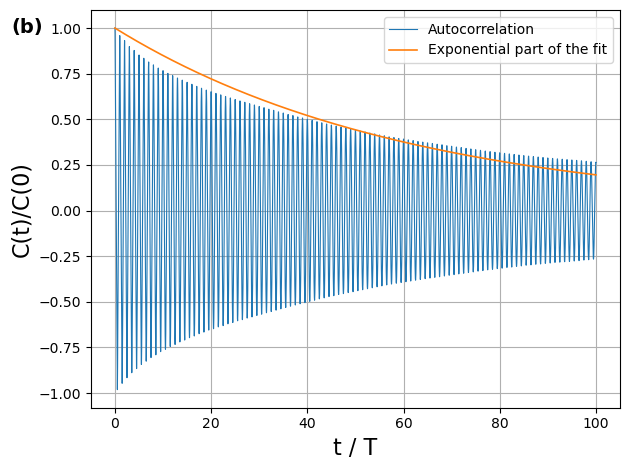}
       \includegraphics[width=0.48\textwidth]{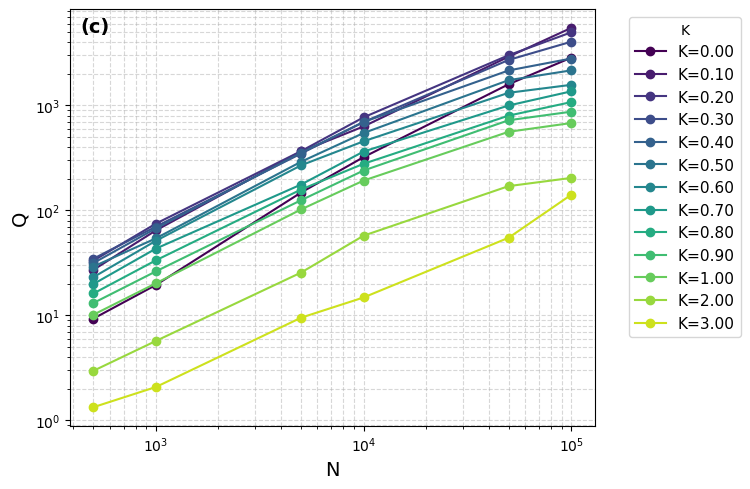}
    \includegraphics[width=0.48\textwidth]{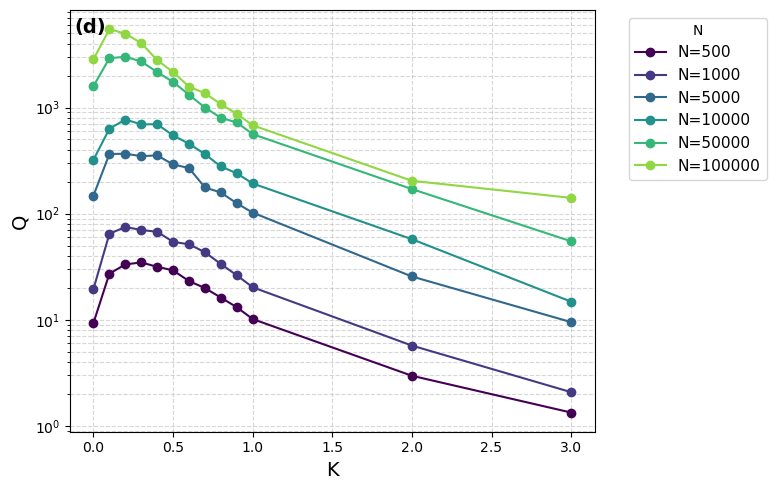}
    \caption{(a,b) Autocorrelation analysis: extraction of diffusion coefficient $D$ and angular frequency $\omega_0$ from $\mathrm{Re}\{C(\tau)\}$, shown for $N=10^4$ at $K=0$ and $K=1$. (c,d) Quality factor $Q$ as a function of $N$ and $K$.}
    \label{fig:Q}
\end{figure}

We then introduce the complex autocorrelation function of the phase $\theta$:
\begin{equation*}
    C(\tau)=\langle e^{i[\theta(t+\tau)-\theta(t)]}\rangle.
\end{equation*}
Its real part exhibits a damped oscillatory behavior, which can be accurately fitted with
\begin{equation*}
    C_{\rm fit}(\tau)=e^{-D\tau}\cos(\omega_0\tau),
\end{equation*}
allowing us to extract the phase diffusion coefficient $D$ and the mean angular frequency $\omega_0$ (the period $2\pi/\omega_0$ is shown in Fig.~\ref{cycles}f). Two examples of the curves $C(\tau)$ are shown in Fig.~\ref{fig:Q}.

Having extracted the oscillation frequency $\omega_0$ and the phase diffusion coefficient $D$, the quality factor is defined as
\begin{equation}
    Q=\frac{\omega_0}{2D}.
\end{equation}
The numerical results for the quality factor are shown in Fig.~\ref{fig:Q}. In (a) the growth with $N$ is shown. For weak coupling ($K \ll 1$), the simulations show that $Q$ grows linearly with the system size $N$, in agreement with the theoretical prediction.  
For moderate to strong coupling ($K \gtrsim 0.5$), the quality factor for large $N$ deviates from the linear behavior, in almost all cases bending down (with the only exception of the case $K=3$) indicating that temporal coherence becomes limited by interaction-induced fluctuations rather than by finite-size noise. We suggest that the results in Fig.~\ref{fig:Q}a could explain apparently contrasting results in the recent literature, i.e. the linear behavior seen in~\cite{Sharma} and sublinear behavior observed in \cite{Maggi2022} could correspond to different intensity of the motor-motor coupling in the two different experiments. In the same Figure it is seen that at constant $N$ the quality factor first increases with $K$, as anticipated by the linear analysis, and then - roughly at the values of $K$ that break the validity of the linear expansion - always decreases with $K$.

\begin{figure}[h!]
    \centering
    \includegraphics[width=\textwidth]{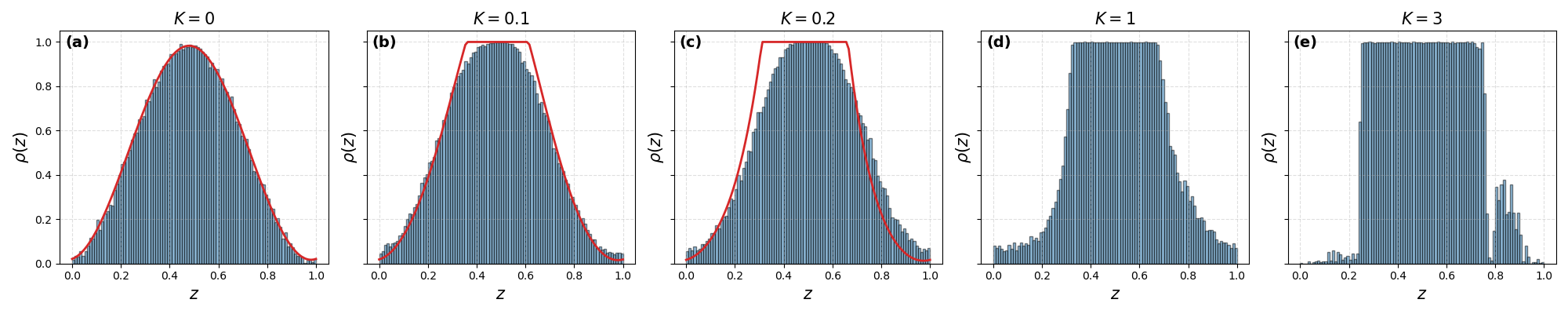}
    \caption{ Density of active motors at the final step, for different values of K. Blue histograms represent numerical simulations. In the weak-coupling regime, they are compared with the analytical prediction (red curve).}
        \label{fig: densità}
\end{figure}

Regarding the density of active motors, the numerical profile is obtained via spatial binning, choosing a number of bins such that each contains a sufficiently large number of motors (here: $100$ bins for $N=100000$).  This empirical density can then be directly compared with the theoretical prediction, as illustrated in Fig.~\ref{fig: densità}.
In the numerical integration of the evolution equations for the Fourier modes,  a truncation had to be imposed. Since the results indicate that the mode amplitudes decay exponentially with increasing index, truncating at a sufficiently high index introduces only negligible errors. We therefore chose $n_{\rm max}=10$.

From the numerical density profiles we can extract several interesting features:
\begin{enumerate}
    \item For $K=0$, the profile is Gaussian and always satisfies $\rho \le 1$ without the need to impose any explicit physical constraint.
    \item For $K>0$, a region with $\rho \ge 1$ appears. This is unphysical, and enforcing the bound $\rho \le 1$ corresponds to a saturated plateau with $\rho=1$, which is precisely what we observe in the simulations.
    \item For small $K$, the theoretical prediction agrees well with the simulations, both in the shape of the density profile and in its temporal evolution. As $K$ increases, theory progressively overestimates the width of the saturated region and its temporal dynamics becomes less accurate. The magnitude of the mean deviation, averaged over space and time, between the theoretical density profile and that obtained from simulations increases with $K$ ($K=0$, $ \overline{|\Delta\rho|}=1.3\%$; $K=0.1$, $\overline{|\Delta\rho|}=5.7\%$; $K=0.2$, $\overline{|\Delta\rho|}=7.4\%$).
    \item For large $K$, where the perturbative theory loses validity, simulations nevertheless show a clear qualitative behavior: strong coupling promotes motor cooperativity to the point that the system splits into two domains, one where all motors are active and one where all are inactive, with only weak modulation induced by the filament motion.
   
\end{enumerate}

\begin{figure}
    \centering
    \includegraphics[width=0.33\textwidth]{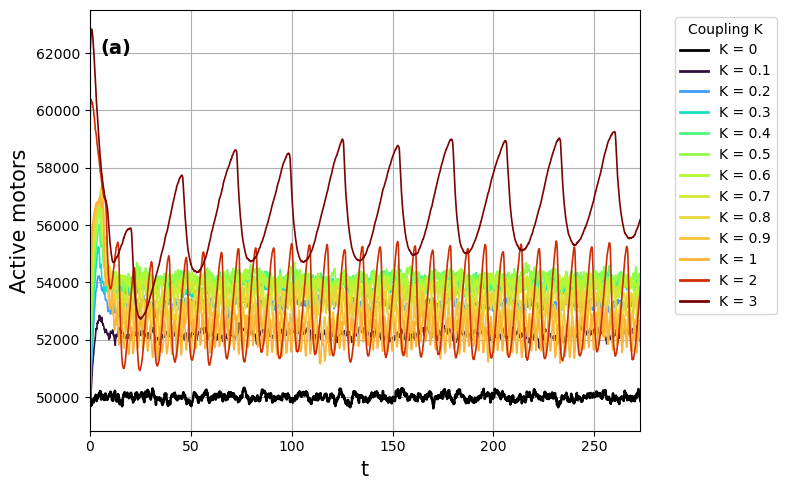}\hfill
    \includegraphics[width=0.33\textwidth]{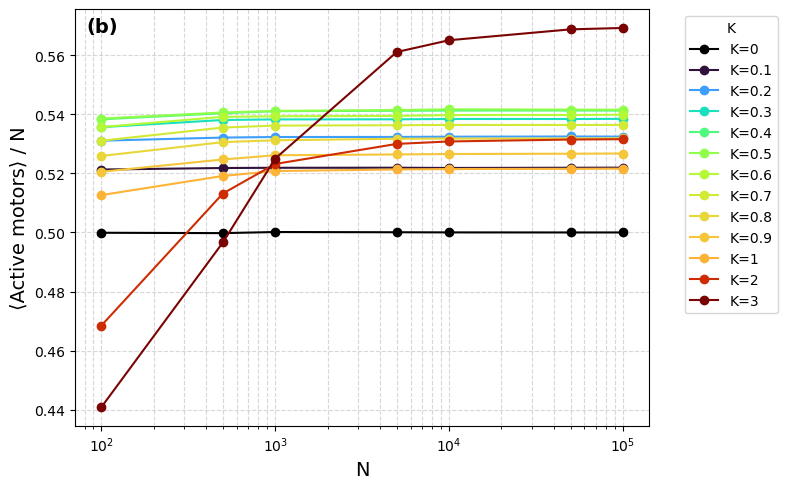}\hfill
    \includegraphics[width=0.33\textwidth]{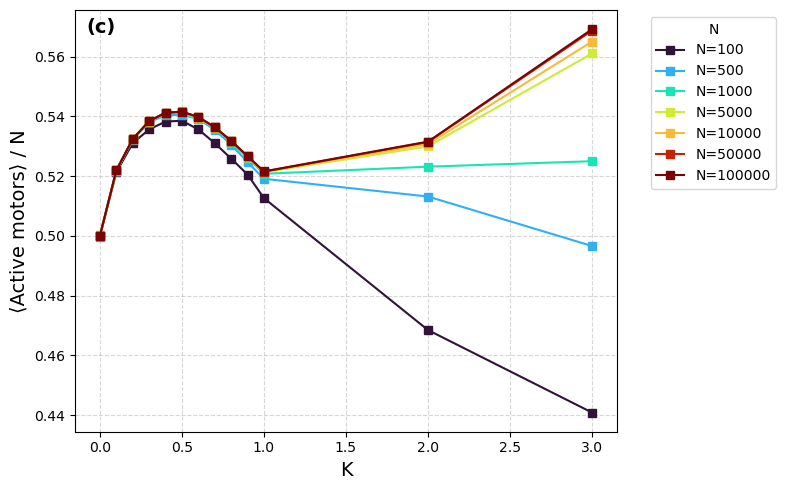}
    \caption{Time evolution of the total number of active motors $N_{\rm act}(t)$.
    The total number of motors is $N=10^5$. (a), average number of active motors as a function of $N$ (b) and $K$ (c). \label{fig:num mot attivi}}
    
\end{figure}

A second interesting observable characterising the effect of $K$ is the total number of active motors. The time behavior of this observable is shown in Fig.~\ref{fig:num mot attivi}, left graph. From the simulations we observe that:
\begin{enumerate}
    \item The time series $N_{\rm act}(t)$ exhibits a pronounced initial peak, corresponding to the spatial reorganization of motors; this peak becomes stronger as $K$ increases.
    \item For small $K$, the average fraction of active motors remains nearly constant around $\eta \approx 0.5$.
    \item For large $K$, the system undergoes a sharp transition toward a regime where the average fraction of active motors is an increasing function of $N$, going from values slightly smaller than $50\%$ to values larger than $50\%$, saturating to $\sim 56 \%$ at very large $N$.
\end{enumerate}

\begin{figure}[h!]
    \centering
    \includegraphics[width=0.5\textwidth]{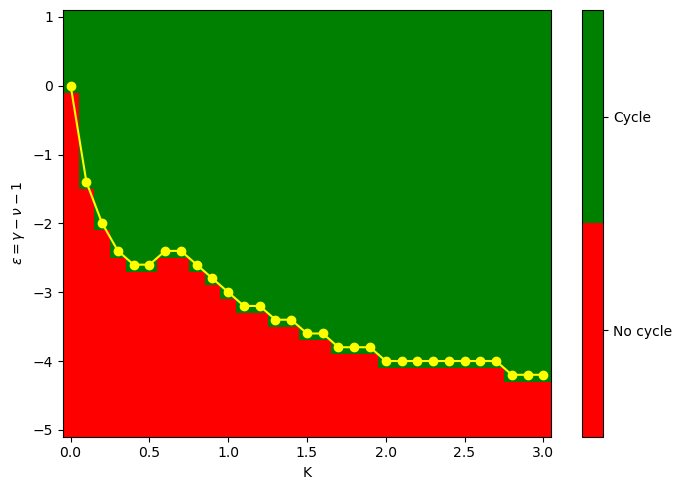}
    \caption{Phase diagram showing the transition from fixed point to limit cycle as a function of $K$. Results are obtained with $N=50000$.}
    \label{fig:phase trans}
\end{figure}

The effect of $K$ is also visible in the stabilization of the system's limit cycle.
In the uncoupled case, the parameter $\epsilon = \gamma - \nu - 1$ defines two distinct zones: $\epsilon > 0$, where a limit cycle exists, and $\epsilon < 0$, where it does not.  
By varying $K$, we can show that the critical value $\epsilon_c$ depends on $K$ and is no longer fixed at $0$.  
This result is shown in Figure~\ref{fig:phase trans}.  
The data were obtained by simulating the system up to a final time $T = 1000$, saving the values of $X$ versus $F$ from $T_0 = 500$ onward to eliminate the initial transient.  
Two conditions must be satisfied to classify the system as exhibiting a limit cycle:

\begin{enumerate}
    \item The ratio between the peak and the noise of the FFT of the signal must be greater than $500$.  
    \item The variance of $X$, expressed in dimensionless units, must be greater than $10^{-4}$. This condition avoids false positives: if the FFT noise is very low, for example when the point oscillates only slightly around a fixed point, the peak-to-noise ratio could become artificially large and incorrectly indicate the presence of a limit cycle.
\end{enumerate}

From this phase diagram, we can see that increasing the coupling $K$ enlarges the region of $\epsilon$ values for which the limit cycle exists. 
This indicates that the coupling promotes collective oscillations, effectively extending the accessible parameter space where a limit cycle can be sustained. 
The critical value $\epsilon_c(K)$ decreases with increasing $K$, initially rapidly for small $K$ and then more slowly for larger values. These observations are consistent with the predictions of Section~\ref{reduced}.

\section{Conclusions} \label{conclusions}

We have studied numerically and analytically a "rigid" model for active flagellar beating, driven by stochastic molecular motors, under the external effect of elastic confinement and viscosity. The
model investigated here is a modification of a model introduced by Julicher and Prost in the 90s~\cite{julicher1995cooperative,Julicher1997}. The modification is a direct coupling between adjacent motors, favouring them to be in the same state. This ingredient, with some technical differences, was proposed recently to understand experimental results concerning phase fluctuations in bull sperm beating~\cite{Maggi2022}. The direct coupling allowed to explain the  low quality factor observed in the experiment. The investigation here includes an analytical approach for small coupling and numerical simulations that span a larger range of coupling constant $K$. The analytical results clearly indicate that increasing $K$ rapidly leads to more and more Fourier modes to become relevant in the spatial distribution of activated motors and that the main relaxation time-scale becomes slower and slower as $K$ increases. This is reflected in an oscillating period that rapidly increases with $K$. When the linear approximation fails (visible only in the simulations), this effect is accompanied by a distortion of the limit cycle shape, with an avalanche-like dynamics: the motors attach and detach in large groups. Correspondingly, the quality factor, which in the small $K$ regime increases with $K$, starts decreasing, coherently with the observations of~\cite{Maggi2022}.

An interesting observation emerging from our study is the similarity between the effect of increasing the motor coupling $K$ and the effect of reducing the elastic confinement coefficient $k$ (or its rescaled counterpart $\nu$). This qualitative similarity includes the increase of the period and the width of the oscillations, as well as the deformation of the limit cycle shape. This observation suggests a possible interpretation of the consequence of large motor couplings: such couplings, by inducing a slower dynamics, lead to a more persistent stochastic active force. The force, acting for longer times in the same direction, stretches more the filament, similarly to what would happen if the elastic force was reduced. Note that when going from $K=0$ to $K=3$ the magnitude of the force goes from $\sim 1$ to $\sim 6$ (see for instance the maximum excursion of $F$ in Fig.~\ref{cycles}e) while the period goes from $\sim 2$ to $\sim 50$. 

In summary, a direct coupling among adjacent molecular motors in elastic flagella has the effect of coordinating larger and larger groups of motors, such that the generated driving force is slower and more effective in producing the oscillation dynamics, eventually {\em breaking} the smooth linear beating and creating an avalanche-like (or stick-slip) dynamics. Interestingly, our results could be a first hint to reconciliate apparently contrasting results in the recent literature~\cite{Maggi2022,Sharma}.

We remark that the conjecture that, in an axoneme, a molecular motor has a direct interaction with its nearest neighboring motors, is supported 
by the observation (in micrographs by scanning electron microscopy, etc.) of nonrandom grouping of dynein states and
by the evidence that interactions between adjacent dyneins
may be inevitable because of the size of dynein arms~\cite{brokaw2002computer,burgess1995rigor,goodenough1982substructure}.

Extending the model to allow for traveling-wave solutions goes through the removal of the elastic confinement~\cite{julicher1995cooperative} and adding a source of spatial symmetry breaking. An asymmetric motor coupling is very likely to break the symmetry inducing a base-to-tip traveling wave. Ideally an investigation of our model with these suggested modifications and a more complex fluid-filament coupling, allowing for real swimming, is a valuable  direction for future work. 

\section*{Acknowledgements}
AP and FF warmly thank M. Baldovin for inspiring discussions, suggestions and a critical reading of the manuscript. AP also thanks F. Cecconi, T. Guerin, D. Lucente, C. Maggi and M. Viale for enlightening discussions. The Authors both thank Reviewer Dr. Philip Bayly for valuable suggestions, including the idea of an asymmetric motor coupling as a realistic ingredient for producing travelling waves and swimming propulsion.

\paragraph{Funding information}
AP  acknowledges funding from the Italian Ministero dell’Università e della Ricerca under the programme PRIN 2022 ("re-ranking of the final lists"), number 2022KWTEB7, cup B53C24006470006.

\newpage

\begin{appendix}

\section{Appendix 1: details of the spatially homogeneous model}
\label{hom}

In this Appendix we give some detail of the spatially homogeneous model, starting with a brief discussion of a simplified Markovian model for a molecular motor, which justifies a velocity-load relation qualitatively similar to what is observed and used in the spatially homogeneous model.

\subsection{A single molecular motor}

One of the simplest stochastic modes for a molecular motor is a Markov jump process with $3$ states~\cite{hwang2018energetic,peliti2021stochastic}. The presence of more than $2$ states allows for breaking detailed balance. The transition rates from state $i$ to state $j$ are denoted by $W_{ij}$. 
Solving the Kolmogorov (Master) equation in the steady state gives the following stationary probabilities to find the motor in state $i$:
\begin{align} \label{steadyprob}
    P_0&=\frac{{W_{10}} ({W_{20}}+{W_{21}})+{W_{20}} {W_{12}}}{{W_{10}} ({W_{20}}+{W_{21}}+{W_{02}})+{W_{20}} ({W_{01}}+{W_{12}})+({W_{01}}+{W_{02}}) ({W_{21}}+{W_{12}})},\\
    P_1&=\frac{{W_{20}} {W_{01}}+{W_{21}} ({W_{01}}+{W_{02}})}{{W_{10}} ({W_{20}}+{W_{21}}+{W_{02}})+{W_{20}} ({W_{01}}+{W_{12}})+({W_{01}}+{W_{02}}) ({W_{21}}+{W_{12}})},\\
    P_2&=1-(P_0+P_1).
\end{align}
The steady state current in the system reads:
\begin{equation} \label{steadycur}
    J=P_0 W_{01}-P_1 W_{10}=P_1 W_{12}-P_2 W_{21}=P_2 W_{20}-P_0 W_{02},
\end{equation}
whose full expression in terms of the rates is not reported here since it is immediate to obtain (by substituting Eq.~\eqref{steadyprob} into~\eqref{steadycur}, long and not particularly transparent. At equilibrium, that is in the absence of energy sources and in the presence of thermal fluctuations only, $w^{ij}_{atp}=0$ for all pairs $ij$, so that $P_i \propto e^{-E_i/(k_B T)}$ (where $E_i$ is the free energy of state $i$ and $k_B T$ is thermal energy at temperature $T$) and $J=0$.

\begin{figure}
    \centering
    \includegraphics[width=0.45\linewidth]{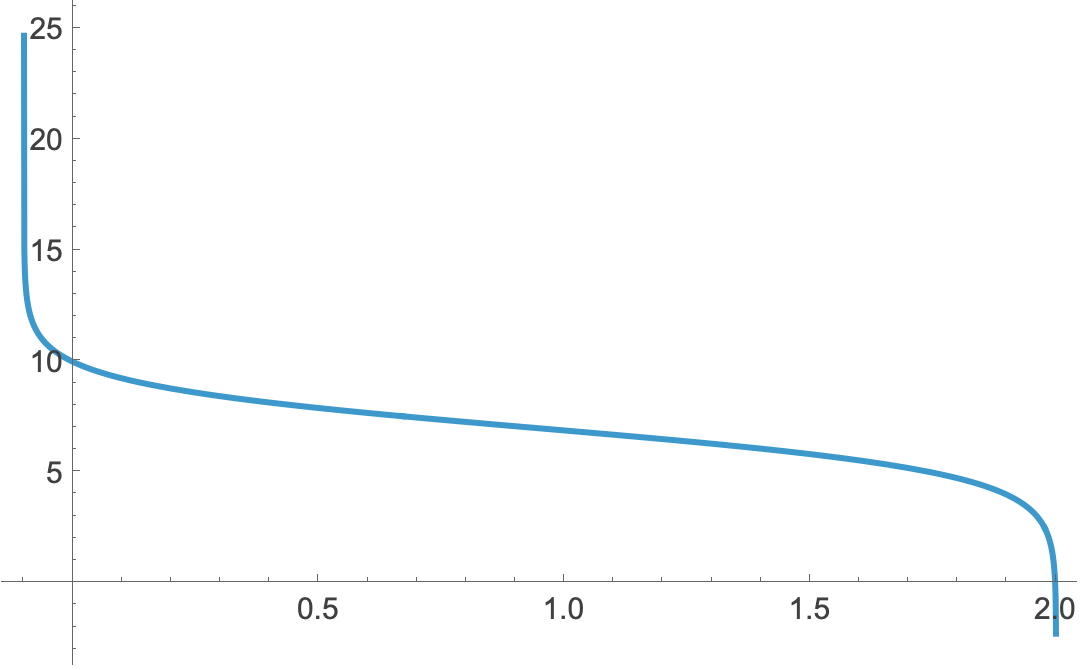}
  \includegraphics[width=0.45\linewidth]{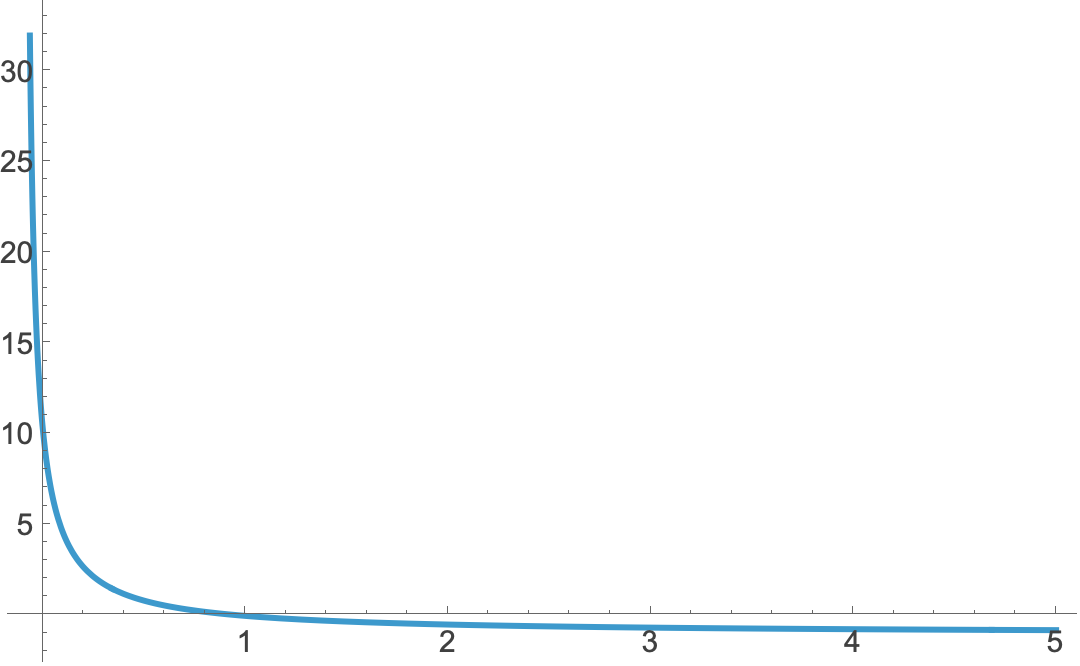} 
    \caption{Load versus velocity characteristic. Left: simple kinesin model.  Right:   model adopted in the paper, similar to what observed in motility-assay experiments, see for instance~\cite{howard2009mechanical} Fig. 2.}
    \label{fig:kinesin}
\end{figure}

Transition rates are characterised by a local detailed balance relation of the kind~\cite{peliti2021stochastic}:
\begin{equation}
    \frac{W_{ij}}{W_{ji}}=e^{-\frac{E_j-E_i-w^{ij}_{atp}}{k_B T}},
\end{equation}
 where $w^{ij}_{atp}=-w^{ji}_{atp}$ is the work energy provided, in the reaction $i \to j$, by ATP hydrolysis. 
The minimal assumption, when ATP is present and the motor can work, is that $w^{ij}_{atp}=0$ for all pairs $ij$ but one, e.g. $i=1$, $j=2$, such that for instance $w_{12}^{atp}=w^*-y$ where $y$ is the external load. In Fig.~\ref{fig:kinesin}, left plot, we show the result of the load-current graph for this model. Actually, motility assays experiments show a slightly different characteristic, for instance see ~\cite{howard2009mechanical} Fig. 2. In the following we discuss the spatially homogeneous model for filaments - i.e. with many motors - using a  model where the load-current characteristic is of the form shown in Fig.~\ref{fig:kinesin} right plot, similar to what seen in motility assays experiments.

\subsection{Motors on a filament: a spatially homogeneous model}

The model, described in~\cite{howard2009mechanical}, considers $N$ motors that can be detached or attached (walking on it) to a "rigid filament", which is at a variable position $X(t)$, while keeping their tails attached to a load (e.g. the spring resistance due to elongation toward a second filament). 
Here there is no information about the position of the motors: the $N$ motors are grouped into $N_a$ active motors and $N-N_a$ inactive motors. The active motors all apply the same velocity-dependent load $y(\dot X)$, independently of where they are located.
The model, based upon experimental evidence through the so-called motility assays~\cite{howard2002mechanics}, assumes that $y(\dot X)$  is a decreasing function with a stall value $y_0=y(v=0)$ and an unloaded velocity $v_0$ such that $y(v_0)=0$.   The filament feels the forces of all attached motors, which are in a number $N_a(t) \le N$, that of a spring attached to a substrate, and the viscous drag (with filament mobility $1/\gamma$) of the surrounding fluid, which neglecting inertia leads to
\begin{equation} \label{eqX}
    \gamma \dot X=-k X+ n_a(t) y(\dot X),  
\end{equation}
where we introduced $n_a(t)=N_a(t)/N$ the fraction of active motors. Here, with respect to the discussion in~\cite{howard2009mechanical}, we decided to consider an active force which is of order $1$ (relatively to the number of motors): this is for coherence with the choice to study the main JP and JPK models with a parameter $\gamma$, in Eq.~\eqref{eq:Xdot_dimless}, which is constant in $N$ (the sum of the effects of the $N$ motors is rescaled by $N$).

Equation~\eqref{eqX} must be complemented by an equation for $N_a(t)$, which comes from the second important assumption of the model that is a detachment rate that depends upon the load  $\tilde\omega_{off}(y)$, and therefore upon the velocity $\omega_{off}(v)=\tilde\omega_{off}(y(v))$ leading to the Kolmogorov equation
\begin{equation} \label{eqNa}
    \dot n_a=-[\omega_{off}(\dot X)+\omega_{on}]n_a+\omega_{on}.
\end{equation}

Fundamental parameters of this model are also
\begin{align}
    y'&=\left.\frac{d y}{d \dot X}\right|_{\dot X=0},\\
    \omega'&=\left.\frac{d \omega_{off}}{d \dot X}\right|_{\dot X=0}=\left.\frac{d \Omega_{off}}{dy}\right|_{y_0} y' = \Omega' y',
\end{align}
where we recall that $y_0=y(0)$. 

Model in Eqs~\eqref{eqX}-\eqref{eqNa} has a fixed point in
\begin{align}
    \dot X&=0,\\
    X^*&=n_a^* \frac{y_0}{k},\\
    n_a^*&=\frac{\omega_{on}}{\overline{\omega}},
\end{align}
where $\overline{\omega}=\omega_{on}+\omega_0$, being $\omega_0=\omega_{off}(\dot X=0)=\Omega_{off}(y_0)$. Linearization around the fixed point by replacing $X(t)=X^*+x(t)$ and $n_a(t)=n_a^*+n(t)$ and keeping only linear terms, gives
\begin{align}
    (\gamma-n_a^* y')\dot x&=- kx + y_0n \label{linx},\\
    \dot n&=-\overline{\omega} n - \omega' n_a^* \dot{x} \label{linn},
\end{align}
whose stability can be studied in two ways, i.e. writing the system as $d/dt (x,n)^T=M (x,n)^T$ and studying the real parts of the eigenvalues of
\begin{equation}
    M = \frac{1}{(\gamma-n_a^* y')} \begin{pmatrix} -k &y_0 \\ \omega' n_a^* k &-[\overline{\omega}(\gamma-n_a^* y') + \omega' n_a^* y_0]\end{pmatrix},
\end{equation}
or as an instructive alternative multiplying Eq.~\eqref{linx} by $\overline{\omega}^{-1}d/dt + 1$ which brings to
\begin{equation}
    \overline{\omega}^{-1} (\gamma-n_a^* y') \ddot x + \dot x = - k\overline{\omega}^{-1} \dot x - k x + y_0(\overline{\omega}^{-1}d/dt + 1) n,
\end{equation}
and the last term can be replaced by looking at Eq.~\eqref{linn} which gives $(\overline{\omega}^{-1}d/dt + 1) n = -\overline{\omega}^{-1}\omega' n_a^* \dot{x}$ and therefore
\begin{equation}
    \overline{\omega}^{-1}(\gamma-n_a^* y') \ddot x + (\gamma-n_a^* y') \dot x = -k\overline{\omega}^{-1} \dot x - k x +y_0(-\overline{\omega}^{-1}\omega' n_a^* \dot{x}),
\end{equation}
that is straightforwardly recast into the following oscillator equation
\begin{equation}
    m_e \ddot x + \gamma_e \dot x +kx=0,
\end{equation}
with
\begin{align}
    m_e = \overline{\omega}^{-1}(\gamma-n_a^* y'), \\
    \gamma_e = \gamma+ \frac{k}{\overline{\omega}}+\left(\frac{y_0\Omega'}{\overline{\omega}}-1\right)n_a^* y',
\end{align}
The effective viscosity can become negative, signaling immediately the loss of stability of the fixed point in favour of oscillations, which of course should be stabilised by non-linear terms not retained here. It is easy to verify that the real part of the eigenvalues of $M$ is the same as that of the second-order oscillator problem and reads $-\gamma_e/(2 m_e)$.

\begin{figure}
    \centering
    \includegraphics[width=0.48\linewidth]{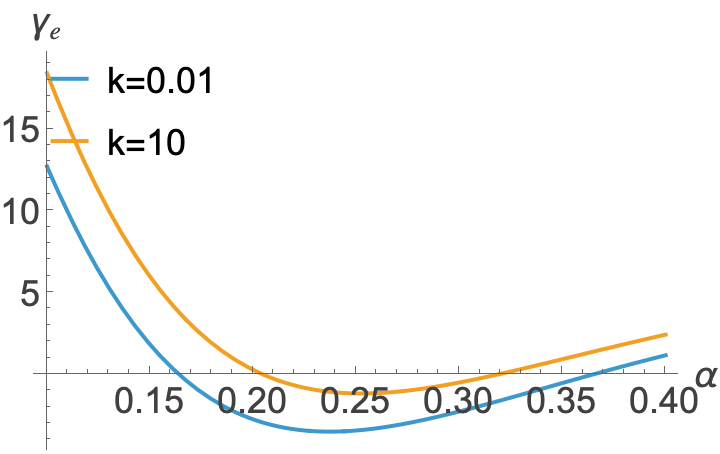}
    \caption{Effective viscosity $\gamma_e$ as a function of $\alpha$ for $y_{min}=-1$,  $y_0=10$, $v_{min}=0.1$, $\Omega_{\infty}=10$, $\Omega_0=0.5$, $\omega_{on}=0.5$, $N=10^3$, $\gamma=1$, with two different values for $k=0.01$ and $k=10$.
    }
    \label{fig:guerin}
\end{figure}

A concrete possibility for the functions $y(\dot X)$ and $\Omega_{off}(y)$ is
\begin{align}
    y(\dot X) &=y_{min}-(y_0-y_{min})\frac{v_{min}}{\dot X-v_{min}} \label{ourchoice},\\
    \Omega_{off}(y) &=\frac{\Omega_{\infty}e^{\alpha y}}{\frac{\Omega_{\infty}}{\Omega_0}-1+e^{\alpha y}}, 
\end{align}
where $y_{min}<0$ is the minimum load (attained at infinite velocity~\footnote{We should  check if in a simple molecular motor model the velocity becomes infinite at a finite or infinite negative load.}), $y_0$ is the stall load (at zero velocity), $v_{min}<0$ is the minimum velocity attained at infinite load, $\Omega_\infty$ is the detachment rate at infinite load (minimum velocity), while $\Omega_0$ is the detachment rate at zero load which is attained when $\dot X=y_0 v_{min}/y_{min}$. With these functions, the function $\omega_{off}(\dot X)$ has a positive minimum $\omega_{min}=\Omega_{off}(y_{min})$ for infinite $\dot X$ (which becomes $0$ if the load model goes to $-\infty$ for infinite velocity), and goes to infinity when $\dot X \to v_{min}<0$ (from above), passing through $\omega_0$. With these choices one gets the following results. In Fig.~\ref{fig:guerin} the effective viscosity $\gamma_e$ is shown as a function of $\alpha$ (which determines the slope of the dissociation-load characteristic), demonstrating a range where it becomes negative.

\begin{figure}[h!]
    \centering
    \includegraphics[width=0.48\linewidth]{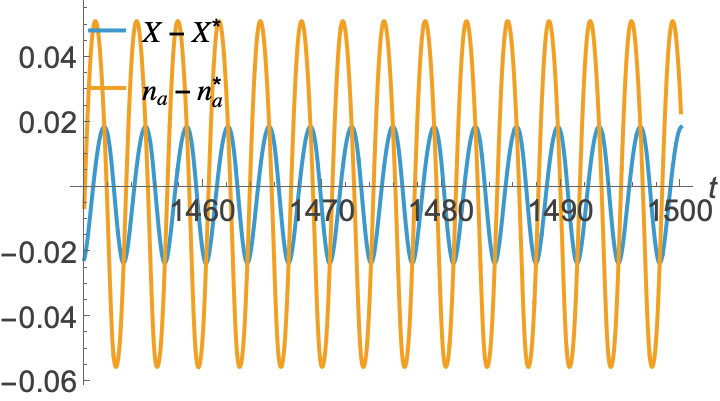}
    \includegraphics[width=0.48\linewidth]{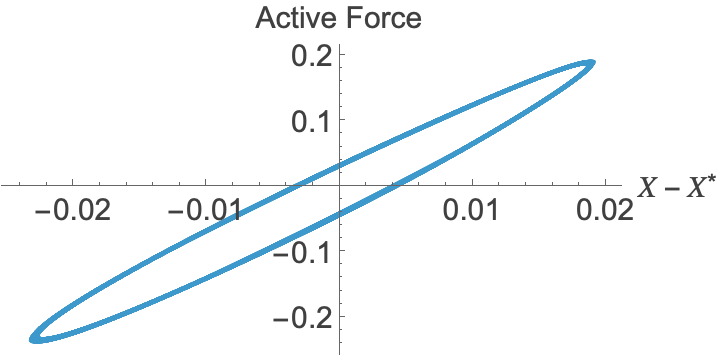}
    \includegraphics[width=0.48\linewidth]{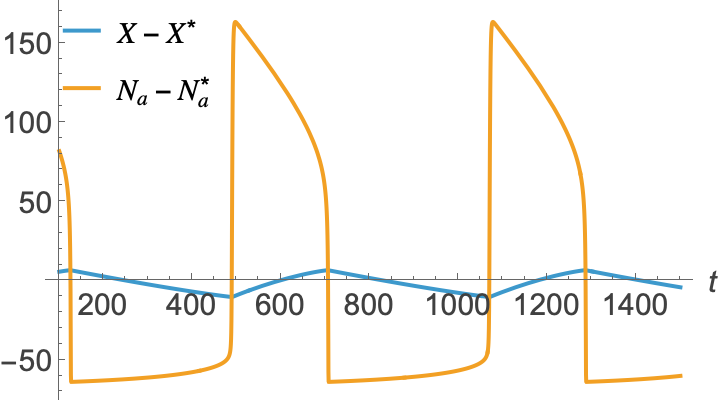}
    \includegraphics[width=0.48\linewidth]{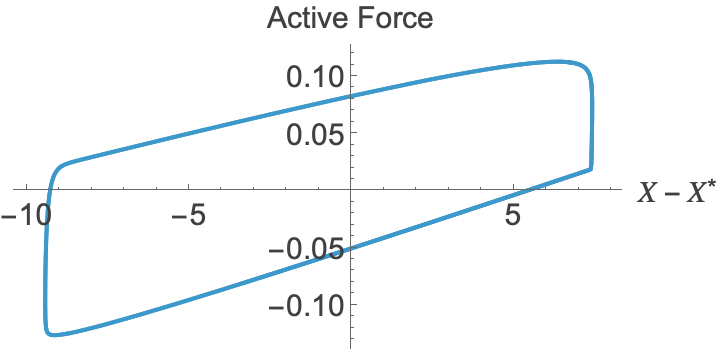}
    \caption{Numerical simulation of the homogeneous model. Left: temporal evolution of the two variables of the model shifted by their respective fixed points: $X(t)-X^*$ and $n_a(t)-n_a^*$ for the first case and $N_a(t)-N_a^*$ (where $N_a^*=n_a^* N$) for the second case. Right: limit cycle in the force-position phase plane of the model, i.e. in abscissa we have the shifted elongation $X(t)-X^*$, in ordinate we have the shifted active force $n_a(t) y[\dot X(t)]-n_a^* y(0)$ . The cycle runs clockwise. Top graphs are with $k=1$, bottom graphs are with $k=0.01$. Other parameters are the same in all graphs: $N=1000$, $\omega_{on}=0.5$, $\alpha=0.25$, $y_{min}=-1$, $v_{min}=-0.1$, $y_0=10$, $\omega_{\infty}=10$, $\omega_0=0.5$. }
    \label{fig:howard}
\end{figure}

An example of temporal evolution of $X(t)-X^*$ and $n_a(t)-n_a^*$, together with the limit cycle in that plane, is shown in Fig.~\ref{fig:howard}, for two different values of $k$. The lesson of this simple model is that when $k$ is large the evolution of the system is smooth with fast periods and the limit cycle is an ellipsis; on the contrary, when $k$ is small the period is much longer and the cycle is hysteretic: there is a moment where the elongation of the filament is minimal (such that the total load is small), at that moment the motors attach very rapidly up to the point where the filament elongation starts to grow, then the motors detach slowly together with the growth of the filament and its load force, this slow growth continues up to the rupture moment where the motor abruptly detach to their minimum value, and remain at that value for all the time needed to the filament to relax back. Then the cycle starts again. Basically there is the presence of two "avalanches" of motor attaching and detaching. The strong increase of the period and the deformation of the limit cycle with larger excursions when $k$ is reduced is strongly reminiscent of what happens in the JPK model when the motor coupling $K$ is increased. 

\newpage
\section{Appendix 2: Analytical derivation of main results}

In this appendix, we present all the analytical results obtained during the study. We show  the results valid for both models (with and without coupling), and those specific to the coupled case (which, in the limit $K = 0$, can also be applied to the uncoupled model).\\
Here there is a table with all the tools  to interpret the results of the chapter.\\

\begin{table}[h!]
\centering
\renewcommand{\arraystretch}{1.3} 
\caption{Summary of notation used in the derivations.}
\begin{tabular}{|p{5cm}|p{8cm}|}
\hline
\textbf{Symbol} & \textbf{Meaning} \\
\hline
$N$ & Total number of motors \\
\hline
$m$ & Number of discrete binding sites \\
\hline
$\epsilon_i = n_i / N$ & Fraction of active motors at site $i$ \\
\hline
$X$ & Filament (flagellum) position \\
\hline
$z$ & Spatial coordinate in lab frame \\
\hline
$\rho(z)$ & Motor density field on interval $[0,\ell]$ \\
\hline
$c_n$, $s_n$ & $\cos\left( \frac{2\pi n z}{\ell} \right)$, $\sin\left( \frac{2\pi n z}{\ell} \right)$ \\
\hline
$c_X$, $s_X$ & $\cos\left( \frac{2\pi X}{\ell} \right)$, $\sin\left( \frac{2\pi X}{\ell} \right)$ \\
\hline
$a_n$, $b_n$ & Fourier coefficients of $\rho(z)$ \\
\hline
$\alpha$, $\eta$ & Parameters of the rate of attachment \\
\hline
$K$ & Coupling strength between motors \\
\hline
$\Omega$ & Characteristic transition rate \\
\hline
$\nu$ & Adimentional friction coefficient \\
\hline
$\xi$ & Drag coefficient \\
\hline
$f(z - X)=\eta-\alpha \cos\left( \frac{2\pi (z-X)}{\ell} \right)$ & Spatially modulated attachment function \\
\hline
$D^{aa}_{mn}, D^{bb}_{mn}$ & Diffusion matrix components for noise analysis \\
\hline
\end{tabular}
\label{tab:notation}
\end{table}
\newpage
\subsection{Rates in the coupled model.} \label{rates}
When couplings between adjacent motors are introduced, modelled by a quadratic potential and the attachment and detachment rates change compared to the uncoupled case.\\
In the model presented in \cite{Maggi2022}, the rates take the form:

\begin{equation}
\omega^{\text{on}}_i = \Omega \left[\eta - \alpha \cos\left( \frac{2\pi (z_i - X(t))}{\ell} \right) \right] e^{ -\Delta U_{\text{bind},i} }, \qquad
\omega^{\text{off}}_i = \Omega - \omega^{\text{on}}_i,
\end{equation}

where \( \Delta U_{\text{bind},i} \) represents the change in the coupling potential associated with the state transition of \( s_i \). This potential is given by:

\begin{equation}
U_{\text{bind},i} = K \left[ (s_i - s_{i+1})^2 + (s_i - s_{i-1})^2 \right],
\end{equation}

with \( K \) quantifying the strength of the interaction between neighbouring motors. The system penalizes configurations where a motor is in a different state than its neighbours, thereby promoting collective coordination.

Since \( s_i \) is a Boolean variable, \( \Delta U_{\text{bind},i} \) can be computed explicitly. Letting \( \bar{s}_i = 1 - s_i \) denote the negation of \( s_i \), and \( \bar{U}_{\text{bind},i} \) the potential computed with \( \bar{s}_i \), we find:

\begin{align}
\nonumber \Delta U_{\text{bind},i} &= \bar{U}_{\text{bind},i} - U_{\text{bind},i}= \\
& \nonumber = K \left[ (\bar{s}_i - s_{i+1})^2 + (\bar{s}_i - s_{i-1})^2 \right] - K \left[ (s_i - s_{i+1})^2 + (s_i - s_{i-1})^2 \right]= \\
& \nonumber= 2K(1 - 2s_i)(1 - s_{i-1} - s_{i+1}).
\end{align}

Noting that the initial state for \( \omega^{\text{on}}_i \) is \( s_i = 0 \), and for \( \omega^{\text{off}}_i \) is \( s_i = 1 \), we can write:

\begin{align}
 \omega^{\text{on}}_i &= \Omega \left[\eta - \alpha \cos\left( \frac{2\pi (z_i - X(t))}{\ell} \right) \right] e^{ -2K(1 - s_{i-1} - s_{i+1})}, \\
\nonumber \omega^{\text{off}}_i &= \Omega - \Omega \left[\eta - \alpha \cos\left( \frac{2\pi (z_i - X(t))}{\ell} \right) \right] e^{ 2K(1 - s_{i-1} - s_{i+1})}.
\end{align}

However, this formulation breaks the uniform rate hypothesis, i.e., \( \omega^{\text{on}}_i + \omega^{\text{off}}_i = \Omega \), which simplifies the problem and enables analytical treatment.\\
From a physical perspective, the exponential term in \( \omega^{\text{on}}_i \) favours activation of \( s_i \) when both neighbours are active (i.e., \( s_{i-1} = s_{i+1} = 1 \)), but in the case of \( \omega^{\text{off}}_i \), if both neighbours are inactive, motor \( i \) is discouraged from switching off. This breaks the desired symmetry in coordination.

Therefore, in this study we consider a slightly modified model compared to \cite{Maggi2022}, in which:

\begin{align}
\omega^{\text{on}}_i &= \Omega \left[\eta - \alpha \cos\left( \frac{2\pi (z_i - X(t))}{\ell} \right) \right] e^{ -2K(1 - s_{i-1} - s_{i+1})}, \\
\nonumber \omega^{\text{off}}_i &= \Omega - \Omega \left[\eta - \alpha \cos\left( \frac{2\pi (z_i - X(t))}{\ell} \right) \right] e^{ -2K(1 - s_{i-1} - s_{i+1})}.
\end{align}

In this way, we recover both the condition \( \omega^{\text{on}}_i + \omega^{\text{off}}_i = \Omega \) and a dynamics that promotes coordination: the motor is more likely to turn on if its neighbours are active, and to turn off if its neighbours are inactive.\\
Another issue that arises when defining the rates is the explicit presence of the \( s_i \) states, which are discrete variables and do not directly appear in the coarse-grained description we are adopting.\\
To overcome this limitation, we can switch to a formulation in terms of average density using a mean-field approach.\\
Let us divide a segment of the filament of length \( \ell \) into \( m \) sites (or "bins") containing a total of \( N \gg m \) motors. Each bin then contains approximately \( N/m \) motors. If the number of motors per site is sufficiently large, we can assume that, except for those at the bin boundaries, the neighbouring motors \( i+1 \) and \( i-1 \) are included in the same bin as motor \( i \).\\
If the density of active motors in site \( j \) is \( \rho_j \), this represents the probability per unit length that a motor is active. Therefore, the expected value of the state \( s_i \) of a motor in that site is:

\begin{equation}
\nonumber \langle s_i \rangle = 1 \cdot \rho_j \ell + 0 \cdot (1 - \rho_j \ell) = \rho_j \ell.
\end{equation}

Assuming that neighbours \( i+1 \) and \( i-1 \) also belong to the same bin, we have:

\[
\langle s_{i+1} \rangle = \langle s_{i-1} \rangle = \langle s_i \rangle = \rho_j \ell.
\]

We can then replace the average value of \( s_i \) in the transition rates, obtaining an average form of the rates that depends on the local density \( \rho_j \):

\begin{align}
\omega^{\text{on}}_j &= \Omega \left[\eta - \alpha \cos\left( \frac{2\pi (z_j - X(t))}{\ell} \right) \right] e^{ -2K(1 - 2\rho_j \ell)}, \\
\nonumber \omega^{\text{off}}_j &= \Omega - \Omega \left[\eta - \alpha \cos\left( \frac{2\pi (z_j - X(t))}{\ell} \right) \right] e^{ -2K(1 - 2\rho_j \ell)}.
\end{align}

Alternatively, in a continuous formulation, replacing the discrete variable \( \rho_j \) with a spatial density field \( \rho(z) \), we obtain:

\begin{align}
\omega^{\text{on}}(z - X(t)) &= \Omega \left[\eta - \alpha \cos\left( \frac{2\pi (z - X(t))}{\ell} \right) \right] e^{ -2K(1 - 2\rho(z)\ell)}, \\
\nonumber \omega^{\text{off}}(z - X(t)) &= \Omega - \Omega \left[\eta - \alpha \cos\left( \frac{2\pi (z - X(t))}{\ell} \right) \right] e^{ -2K(1 - 2\rho(z)\ell)}.
\end{align}

\subsection{Derivation of the Fokker-Planck Equation}

Regardless of the specific form of the attachment and detachment rates, a Fokker-Planck equation can be derived, which therefore holds in both the coupled and uncoupled cases.

First, we define the intensive quantity $\epsilon_i = \frac{n_i}{N}$.  
We can define the transition rates for a master equation as:

\begin{equation}
   \nonumber W(n_i \rightarrow n_i +1) = n_i \, \omega_{\text{off}}(z_i - X),  ~~~~~~~~~~~~ W(n_i \rightarrow n_i - 1) = \left( \frac{N}{m} - n_i  \right) \omega_{\text{on}}(z_i - X).
\end{equation}

In both cases, the meaning is clear: there is a combinatorial factor counting how many ways we can choose a motor among the active ones at site $i$ ($n_i$) or the inactive ones ($\frac{N}{m} - n_i$), corresponding to detachment or attachment, respectively.

Expressing them in terms of $\epsilon_i$:

\begin{equation}
   \nonumber \hat{W}(\epsilon_i \rightarrow \epsilon_i + 1) = N \epsilon_i \, \omega_{\text{off}}(z_i - X),  ~~~~~~~~~~~~ \hat{W}(\epsilon_i \rightarrow \epsilon_i - 1) = \left( \frac{N}{m} - N \epsilon_i  \right) \omega_{\text{on}}(z_i - X),
\end{equation}

and denoting by $w(X\rightarrow X')$ the transition rates from $X$ to $X'$, we can write a master equation for the joint probability of the flagellum position and motor states:

\begin{align}
\nonumber \frac{\partial P(X, \{\epsilon\}, t)}{\partial t} &= \int w(X \rightarrow X') P(X', \{\epsilon\}, t) \, dX' 
- \int w(X \rightarrow X') P(X, \{\epsilon\}, t) \, dX' + \\
&
\nonumber + \sum_i P\left(X, \epsilon_i + \frac{1}{N}, t\right) \hat{W}\left(\epsilon_i + \frac{1}{N} \rightarrow \epsilon_i\right) + P\left(X, \epsilon_i - \frac{1}{N}, t\right) \hat{W}\left(\epsilon_i - \frac{1}{N} \rightarrow \epsilon_i\right) + \\
&
\nonumber  - \sum_i P(X, \epsilon_i, t) \left[ 
\hat{W}\left(\epsilon_i \rightarrow \epsilon_i - \frac{1}{N}\right) 
+ \hat{W}\left(\epsilon_i \rightarrow \epsilon_i + \frac{1}{N}\right) 
\right].
\end{align}

Following Van Kampen~\cite{kampen1961power,van1992stochastic}, we can express the rates as functions defined from an initial point $a_i$ and a jump $r$, in particular: $\hat{W}(a_i \rightarrow a_i + r) = \bar{W}(a_i; r)$.

In the master equation, we therefore have terms of the type:

\begin{equation}
    \nonumber P(X,\epsilon_i \pm \frac{1}{N},t)\bar{W}(\epsilon_i \pm \frac{1}{N}; \mp\frac{1}{N}) .
\end{equation}

We now expand them for $N \gg 1$ up to order $\frac{1}{N^2}$:

\begin{align}
    \nonumber 
    &P(X,\epsilon_i \pm \frac{1}{N},t)\bar{W}(\epsilon_i \pm \frac{1}{N}; \mp\frac{1}{N}) 
    \simeq P(X,\epsilon_i,t)\bar{W}(\epsilon_i; \mp\frac{1}{N}) \\
    \nonumber 
    &\quad \pm \frac{1}{N} \frac{\partial}{\partial \epsilon_i}\left[ P(X,\epsilon_i,t)\bar{W}(\epsilon_i; \mp\frac{1}{N}) \right]
    + \frac{1}{2N^2} \frac{\partial^2}{\partial \epsilon_i^2}\left[ P(X,\epsilon_i,t)\bar{W}(\epsilon_i; \mp\frac{1}{N}) \right].
\end{align}

We can then rewrite the master equation as a Fokker-Planck equation for the variable $\epsilon_i$:

\begin{align}
\nonumber \frac{\partial P(X, \{\epsilon\}, t)}{\partial t} &= \int w(X \rightarrow X') P(X', \{\epsilon\}, t) \, dX' 
- \int w(X \rightarrow X') P(X, \{\epsilon\}, t) \, dX' + \\
& \nonumber  +\frac{1}{N} \frac{\partial}{\partial \epsilon_i}\left[ P(X,\epsilon_i,t)\bar{W}(\epsilon_i; -\frac{1}{N}) - P(X,\epsilon_i,t)\bar{W}(\epsilon_i; +\frac{1}{N}) \right]+ \\
&  \nonumber + \frac{1}{2N^2} \frac{\partial^2}{\partial \epsilon_i^2}\left[ P(X,\epsilon_i,t)\bar{W}(\epsilon_i; -\frac{1}{N}) + P(X,\epsilon_i,t)\bar{W}(\epsilon_i; +\frac{1}{N}) \right].
\end{align}

For $X$, we have a deterministic equation coming from the balance of forces:

\begin{equation}
 \nonumber   N \xi \dot{X} = \sum_i W'(z_i - X)n_i - KX.
\end{equation}

From here, we extract $\dot{X}$ and plug it into the Fokker-Planck equation.  
By also inserting the explicit expressions for the $\bar{W}$ rates, we get:

\begin{align}
\frac{\partial P}{\partial t} =
&- \frac{\partial}{\partial X} \left[ \left(\sum_{i=1}^{m} \frac{W(z_i - X) \epsilon_i}{\xi } - \nu X\right) P \right]  +\notag \\
&\quad + \frac{1}{N}\sum_{i=1}^m \frac{\partial}{\partial \epsilon_i} \left[ \left( \omega_{\text{off}}(z_i - X) N \epsilon_i - \omega_{\text{on}}(z_i - X)\left( \frac{N}{m} - N \epsilon_i \right) \right) P \right] +\notag \\
&\quad + \frac{1}{2N^2} \sum_{i=1}^m \frac{\partial^2}{\partial  \epsilon_i^2} \left[ \left( \omega_{\text{off}}(z_i - X) N \epsilon_i + \omega_{\text{on}}(z_i - X)\left( \frac{N}{m} - N \epsilon_i \right) \right) P \right].
\end{align}

We thus recover the same equation obtained in the uncoupled  \cite{guerin2011bidirectional}, identifying $n_i=N\epsilon_i$
We can then follow the same procedures used by the uncoupled models to derive a continuous description of the problem, obtaining the functional Fokker-Planck equation for $P([\rho(z)], X, t)$:

\begin{align} \label{eq:FP funzionale}
\frac{\partial P}{\partial t} &= - \frac{\partial}{\partial X} \left[ v(X, [\rho]) P \right] 
+ \int_0^l dz \, \frac{\delta}{\delta \rho(z)} \left[ A(z, X, \rho(z)) P \right] +\notag \\
&\quad + \frac{1}{2N} \int_0^l dz \int_0^l dy \, \delta(z - y) \frac{\delta^2}{\delta \rho(z) \delta \rho(y)} \left[ B(z, X, \rho(z)) P \right],
\end{align}

where:

\begin{align}
v(X, [\rho]) &= \frac{1}{\xi} \int_0^l dz \, W'(z - X) \rho(z) - \nu X,\\
A(z, X, \rho) &= \omega_{\text{off}}(z - X) \rho(z) - \omega_{\text{on}}(z - X) \left[ \frac{1}{l} - \rho(z) \right], \\
B(z, X, \rho) &= \omega_{\text{off}}(z - X) \rho(z) + \omega_{\text{on}}(z - X) \left[ \frac{1}{l} - \rho(z) \right].
\end{align}

\subsection{Equations for the Fourier modes}
Equation \ref{eq:FP funzionale} provides a deterministic equation for the evolution of the filament position and a stochastic equation for the evolution of the density field $\rho(z)$.\\
However, the stochastic equation includes a noise term that scales as $\frac{1}{N}$ and can therefore be neglected in the thermodynamic limit.\\
The deterministic equation for the evolution of the density then becomes:
\begin{equation}
    \dot{\rho}(z) = -A(z, X, \rho) = -\omega_{\text{off}}(z - X) \rho(z) + \omega_{\text{on}}(z - X) \left[ \frac{1}{\ell} - \rho(z) \right].
\end{equation}

In our case, we impose the uniform rate hypothesis, so the equation simplifies to:
\begin{equation}
    \nonumber \dot{\rho}(z) = -\Omega \rho(z) + \omega_{\text{on}}(z - X(t)) \frac{1}{\ell}.
\end{equation}

This equation still remains difficult to solve, mainly because, in the presence of coupling, $\omega_{\text{on}}$ depends on the density through an exponential.\\
If one wanted to follow the same analytical approach as Guérin et al. \cite{guerin2011bidirectional}, decomposing the density in a Fourier series to obtain an equation for each mode, the expansion of a function like $e^\rho$ would lead to convolution terms involving all the Fourier coefficients.\\
This would result in a purely formal solution, difficult to handle both analytically and interpretatively.\\
In light of this, we choose to consider weak coupling between motors ($K \ll 1$).\\
Under this approximation, $\omega_{\text{on}}$ can be expanded as follows:
\begin{equation}
    \omega^{\text{on}}_i = \Omega f(z - X) e^{-2K(1 - 2\rho(z)\ell)} \approx \Omega f(z - X)(1 - 2K(1 - 2\rho(z)\ell)) = \Omega(1 - 2K)f(z - X) + 4K\Omega f(z - X) \rho(z),
\end{equation}

where, for brevity, we define:
\begin{equation}
    \nonumber f(z - X) = \left[ \eta - \alpha \cos\left( \frac{2\pi (z - X(t))}{\ell} \right) \right] = \eta - \alpha \cos\left( \frac{2\pi X(t)}{\ell} \right) \cos\left( \frac{2\pi z}{\ell} \right) - \alpha \sin\left( \frac{2\pi X(t)}{\ell} \right) \sin\left( \frac{2\pi z}{\ell} \right)
\end{equation}
Note that the uniform rate hypothesis is crucial for obtaining a linear problem in $\rho$. Without it, even at first order in $K$, the problem would involve quadratic terms in $\rho$, and the equations would couple all Fourier coefficients through convolution terms, similarly to what happens with the function $e^\rho$ mentioned earlier.
We now introduce a few abbreviations to lighten the notation:
\begin{align}
    \nonumber c_n = \cos\left( \frac{2\pi n z}{\ell} \right), \quad s_n = \sin\left( \frac{2\pi n z}{\ell} \right), \\
    \nonumber c_X = \cos\left( \frac{2\pi X}{\ell} \right), \quad s_X = \sin\left( \frac{2\pi X}{\ell} \right),
\end{align}
so we can rewrite:
\begin{equation}
    \nonumber f(z - X) = \eta - \alpha c_X c_1 - \alpha s_X s_1.
\end{equation}

The drift term $A(z, X, \rho)$ thus becomes:
\begin{equation}
    A(z, X, \rho) = \Omega \rho(z) - \frac{\Omega}{\ell}(1 - 2K)f(z - X) - 4\Omega K f(z - X) \rho(z).
\end{equation}

At this point, we can expand the function $\rho(z)$ in a Fourier series, since $z \in [0, \ell]$, multiplying by a factor $\alpha$ :
\begin{equation}
    \nonumber \rho(z) = \alpha a_0 + \alpha \sum_{n=1}^{\infty} a_n \cos\left( \frac{2\pi n z}{\ell} \right) + b_n \sin\left( \frac{2\pi n z}{\ell} \right) = \alpha a_0 + \alpha \sum_{n=1}^{\infty} a_n c_n + b_n s_n.
\end{equation}

However, the term $A(z, X, \rho)$ also includes the product $f(z - X)\rho(z)$ and, since $f$ contains sines and cosines, this product leads to coupling between Fourier coefficients of different orders. We therefore analyze this term separately.

\begin{equation}
    \nonumber
    f(z-x)\rho(z)
    =\bigl(\eta-\alpha\,c_X c_1-\alpha\,s_X s_1\bigr)\rho(z)
    =\eta\,\rho(z)-\alpha\,c_X c_1\,\rho(z)-\alpha\,s_X s_1\,\rho(z).
\end{equation}

Using the following trigonometric identities:
\[
\cos A \cos B = \frac12\Bigl[\cos(A+B)+\cos(A-B)\Bigr],\qquad
\sin A \sin B = \frac12\Bigl[\cos(A-B)-\cos(A+B)\Bigr],
\]
\[
\cos A \sin B = \frac12\Bigl[\sin(A+B)-\sin(A-B)\Bigr],
\]
we find
\begin{align}
   \nonumber
   c_1\rho(z)
   &=c_1\!\Bigl(\alpha a_0+\alpha\sum_{n=1}^{\infty}a_n c_n+b_n s_n\Bigr) =\alpha a_0 c_1+\alpha\sum_{n=1}^{\infty}a_n c_1 c_n+b_n s_1 s_n =\\
   \nonumber
   &=\alpha a_0 c_1
     +\alpha\sum_{n=1}^{\infty}\frac{a_n}{2}\bigl(c_{n+1}+c_{n-1}\bigr)
     +\frac{\alpha b_n}{2}\bigl(s_{n+1}+s_{n-1}\bigr),
\end{align}
and
\begin{align}
   \nonumber
   s_1\rho(z)
   &=s_1\!\Bigl(\alpha a_0+\alpha\sum_{n=1}^{\infty}a_n c_n+b_n s_n\Bigr)=\alpha a_0 s_1+\alpha\sum_{n=1}^{\infty}a_n s_1 c_n+b_n s_1 s_n =\\
   \nonumber
   &=\alpha a_0 s_1
     +\alpha\sum_{n=1}^{\infty}\frac{a_n}{2}\bigl(s_{n+1}+s_{n-1}\bigr)
     +\frac{\alpha b_n}{2}\bigl(c_{n-1}-c_{n+1}\bigr).
\end{align}

We now rescale the sums so that only $c_n$ and $s_n$ terms remain:
\begin{align*}
    \sum_{n=1}^{\infty}a_n c_{n+1}&\xrightarrow{\,n+1\to n\,}\sum_{n=2}^{\infty}a_{\,n-1}c_n,
    &\sum_{n=1}^{\infty}a_n c_{n-1}&\xrightarrow{\,n-1\to n\,}\sum_{n=0}^{\infty}a_{\,n+1}c_n,\\[6pt]
    \sum_{n=1}^{\infty}a_n s_{n+1}&\xrightarrow{\,n+1\to n\,}\sum_{n=2}^{\infty}a_{\,n-1}s_n,
    &\sum_{n=1}^{\infty}a_n s_{n-1}&\xrightarrow{\,n-1\to n\,}\sum_{n=0}^{\infty}a_{\,n+1}s_n
    =\sum_{n=1}^{\infty}a_{\,n+1}s_n.
\end{align*}

Hence,
\begin{align}
   \nonumber
   c_1\rho(z)=\alpha a_0 c_1
   +\frac{\alpha}{2}\!\Bigl[
        a_1+a_2 c_1
        +\sum_{n=2}^{\infty}(a_{\,n+1}+a_{\,n-1})c_n
        +b_2 s_1
        +\sum_{n=2}^{\infty}(b_{\,n+1}+b_{\,n-1})s_n
   \Bigr],
\end{align}
\begin{align}
   \nonumber
   s_1\rho(z)=\alpha a_0 s_1
   +\frac{\alpha}{2}\!\Bigl[
        a_2 s_1
        +\sum_{n=2}^{\infty}(a_{\,n+1}+a_{\,n-1})s_n
        +b_1+b_2 c_1
        +\sum_{n=2}^{\infty}(b_{\,n+1}-b_{\,n-1})c_n
   \Bigr].
\end{align}

Therefore,
\begin{align*}
f(z-X)&\rho(z)=\eta\,\rho(z)
-\frac{\alpha^{\,2}}{2}\Bigl\{
c_X\Bigl[
     2a_0 c_1+a_0 c_1 a_1+a_2 c_1
     +\sum_{n=2}^{\infty}(a_{\,n+1}+a_{\,n-1})c_n
     +b_2 s_1
     +\sum_{n=2}^{\infty}(b_{\,n+1}+b_{\,n-1})s_n
\Bigr] +\\
&+s_X\Bigl[
     2a_0 s_1+a_2 s_1
     +\sum_{n=2}^{\infty}(a_{\,n+1}+a_{\,n-1})s_n
     +b_1+b_2 c_1
     +\sum_{n=2}^{\infty}(b_{\,n+1}-b_{\,n-1})c_n
\Bigr]\Bigr\} =\\[6pt]
&=\eta\,\rho(z)
-\frac{\alpha^{\,2}}{2}\Bigl\{
c_X a_1+s_X b_1
+\bigl(2a_0+a_2\bigr)c_X c_1
+\bigl(2a_0+a_2\bigr)s_X s_1
+b_2(c_X s_1+s_X c_1)+ \\
&+\sum_{n=2}^{\infty}\bigl[(a_{\,n+1}+a_{\,n-1})c_X+(b_{\,n+1}-b_{\,n-1})s_X\bigr]c_n +\sum_{n=2}^{\infty}\bigl[(a_{\,n+1}+a_{\,n-1})s_X+(b_{\,n+1}+b_{\,n-1})c_X\bigr]s_n
\Bigr\}.
\end{align*}

Thus, I can rewrite the drift term $ A $ as follows:
\begin{align*}
A(z, X, \rho) &= \Omega \rho(z) - \frac{\Omega}{\ell}(1 - 2K)f(z - X) - 4\Omega K f(z - X) \rho(z) =\\
&= \Omega(1 - 4\eta K)\rho(z) - \frac{\Omega}{\ell}(1 - 2K)f(z - X) - 4\Omega K\left(-\frac{\alpha^2}{2}\right)\left\{ \ldots \right\}= \\
&= \Omega(1 - 4\eta K)\sum_{n=0}^{\infty}(a_n c_n + b_n s_n) 
- \frac{\Omega}{\ell}(1 - 2K)\eta 
+ \frac{\Omega}{\ell}(1 - 2K)\alpha c_1 c_X 
+ \frac{\Omega}{\ell}(1 - 2K)\alpha s_1 s_X+ \\
&\quad + 2\Omega K \alpha^2 \bigg\{
c_X a_1 + s_X b_1
+ (2a_0 + a_2)c_X c_1
+ (2a_0 + a_2)s_X s_1
+ b_2(c_X s_1 + s_X c_1)+  \\
&\quad +  \sum_{n=2}^{\infty} \left[(a_{n+1} + a_{n-1}) c_X + (b_{n+1} - b_{n-1}) s_X \right] c_n 
+ \sum_{n=2}^{\infty} \left[(a_{n+1} + a_{n-1}) s_X + (b_{n+1} + b_{n-1}) c_X \right] s_n
\bigg\}.
\end{align*}

Using the fact that
\begin{equation}
    \nonumber \dot\rho(z)=\alpha\sum_{n=0}^{\infty}(\dot a_n c_n + \dot b_n s_n) .
\end{equation}
and that sine and cosine form an orthonormal basis in Fourier series expansion, we can view the equation $ \dot \rho = -A $ as an equation for the time evolution of the individual Fourier coefficients, leading to the following set of equations:
\begin{align*}
     \dot a_0 &= -\left[ \Omega(1-4\eta K)a_0 - \frac{\Omega}{\ell}(1-2K)\frac{\eta}{\alpha} +2\Omega K \alpha(c_Xa_1+s_Xb_1)\right],\\
     \dot a_1 &= -\left[\Omega(1-4\eta K)a_1+\frac{\Omega}{\ell}(1-2K)c_X+2\Omega K \alpha(2c_Xa_0+c_Xa_2+s_Xb_2)\right], \\
     \dot b_1 &= -\left[\Omega(1-4\eta K)b_1+\frac{\Omega}{\ell}(1-2K)s_X+2\Omega K \alpha(2s_Xa_0+s_Xa_2+c_Xb_2)\right],\\
     & \text{for } n \geq 2,\\
     \dot a_n &= -\left[\Omega(1-4\eta K)a_n+2\Omega K \alpha\left((a_{n+1}+a_{n-1})c_X+((b_{n+1}-b_{n-1})s_X\right)\right],\\
     \dot b_n &= -\left[\Omega(1-4\eta K)b_n+2\Omega K \alpha\left((a_{n+1}+a_{n-1})s_X+((b_{n+1}+b_{n-1})c_X\right)\right].
\end{align*}

To this system of equations, we must also add the equation governing the time evolution of the filament's position.\\
From the Fokker-Planck equation \ref{eq:FP funzionale}, we can write:
\begin{equation*}
     \dot X=v(X, [\rho]) = \frac{1}{\xi} \int_0^l dz \, W'(z - X) \rho(z) - \nu X .
\end{equation*}
We can again connect this to the Fourier coefficients of $ \rho $:
\begin{align*}
    \dot X=\frac{1}{\xi} \int_0^l &dz \, W'(z - X) \rho(z) - \nu X =\frac{2\pi U}{\ell \xi}\int_0^l dz \sin\left( \frac{2 \pi (z-X)}{\ell}\right)\rho(z)-\nu X=\\
    &=\frac{2\pi U}{\ell \xi}\int_0^l dz (s_1c_X-c_1s_X)\alpha\sum_{n=0}^\infty(a_nc_n+b_ns_n)-\nu X.
\end{align*}

Using the fact that the trigonometric functions form an orthonormal basis, the only nonzero integrals are:
\begin{equation*}
    \int_0^l dz\, s_1^2 = \int_0^l dz\, c_1^2 = \frac{\ell}{2}.
\end{equation*}

Thus,
\begin{equation*}
    \dot X = \frac{\pi U \alpha}{\xi}(b_1c_X - a_1s_X) - \nu X 
    = \frac{\gamma}{2 \pi} \ell^2\Omega(b_1c_X - a_1s_X) - \nu X.
\end{equation*}

Therefore, the full system of equations governing the time evolution of the system is:
\begin{center}
\begin{align}
     \dot X &= \frac{\gamma}{2 \pi} \ell^2\Omega(b_1c_X - a_1s_X) - \nu X, \\
 \nonumber    \dot a_0 &= -\left[ \Omega(1-4\eta K)a_0 - \frac{\Omega}{\ell}(1-2K)\frac{\eta}{\alpha} +2\Omega K \alpha(c_Xa_1+s_Xb_1)\right],\\
 \nonumber    \dot a_1 &= -\left[\Omega(1-4\eta K)a_1+\frac{\Omega}{\ell}(1-2K)c_X+2\Omega K \alpha(2c_Xa_0+c_Xa_2+s_Xb_2)\right], \\
 \nonumber    \dot b_1 &= -\left[\Omega(1-4\eta K)b_1+\frac{\Omega}{\ell}(1-2K)s_X+2\Omega K \alpha(2s_Xa_0+s_Xa_2+c_Xb_2)\right],\\
 &\nonumber \text{for } n \geq 2,\\
  \nonumber   \dot a_n &= -\left[\Omega(1-4\eta K)a_n+2\Omega K \alpha\left((a_{n+1}+a_{n-1})c_X+((b_{n+1}-b_{n-1})s_X)\right)\right],\\
  \nonumber   \dot b_n &= -\left[\Omega(1-4\eta K)b_n+2\Omega K \alpha\left((a_{n+1}+a_{n-1})s_X+((b_{n+1}+b_{n-1})c_X)\right)\right].
\end{align}
\end{center}

The system of equations described above has several interesting structural properties. \\
First of all, the equations for the harmonic coefficients show a local coupling structure: each coefficient is coupled only to its nearest neighbours. This type of "nearest-neighbour" interaction is typical of systems with discrete translational symmetry, such as chains of oscillators or tight-binding models. \\
All coefficients $a_n$, $b_n$ (including $a_0$, $a_1$, $b_1$) are subject to a dissipative term proportional to themselves, with prefactor $\Omega(1 - 4\eta K)$. This term plays a key role in the stability of the system: if the prefactor becomes negative, i.e., if $K > \frac{1}{4\eta}$, dissipation turns into exponential growth, leading to linear instability. This sets an upper bound on $K$, which is necessary to ensure stability, and marks the point where the small-$K$ approximation ($K \ll 1$) clearly breaks down. \\
The equation for $\dot X$ is also important: the dynamics of the collective variable $X$ is driven by an active force proportional to a linear combination of $a_1$ and $b_1$, weighted respectively by $-s_X$ and $c_X$. This means that the feedback between the internal configuration (encoded in the harmonics) and the collective motion is mediated by the first spectral components. \\
Finally, the equation for $\dot a_0$ contains a constant term (independent of $a_n$, $b_n$), which acts as a constant driving force, scaled by $1/\ell$ and the parameters $\eta$, $\alpha$, and $K$. This term can potentially push the system toward a non-trivial steady-state configuration, even in the absence of higher-order harmonics. \\
The system can be simplified by introducing dimensionless variables, in which  we set $\ell=\Omega=1$, so it becomes:

\begin{center}
\begin{align}
     \label{eq:sistema ab}\dot X &= \frac{\gamma}{2 \pi} (b_1c_X - a_1s_X) - \nu X ,\\
     \nonumber\dot a_0 &= -\left[ (1-4\eta K)a_0 - (1-2K)\frac{\eta}{\alpha} +2 K \alpha(c_Xa_1+s_Xb_1)\right],\\
     \nonumber\dot a_1 &= -\left[(1-4\eta K)a_1+(1-2K)c_X+2 K \alpha(2c_Xa_0+c_Xa_2+s_Xb_2)\right], \\
     \nonumber\dot b_1 &= -\left[(1-4\eta K)b_1+(1-2K)s_X+2 K \alpha(2s_Xa_0+s_Xa_2+c_Xb_2)\right],\\
     \nonumber\dot a_n &= -\left[(1-4\eta K)a_n+2 K \alpha\left((a_{n+1}+a_{n-1})c_X+((b_{n+1}-b_{n-1})s_X)\right)\right],\\
     \nonumber\dot b_n &= -\left[(1-4\eta K)b_n+2 K \alpha\left((a_{n+1}+a_{n-1})s_X+((b_{n+1}+b_{n-1})c_X)\right)\right].
\end{align}
\end{center}

\subsection{A perturbative approach}

To obtain Eqs.~\ref{eq:sistema ab}, a perturbative approach has already been used, expanding the exponential in the rates so as to make the problem linear.  
The resulting equations are therefore valid in the limit $K \ll 1$ and to first order in $K$.

At this order we can expand the coefficients as
\[
a_n \simeq a_n^{(0)} + K\,a_n^{(1)}, 
\qquad 
b_n \simeq b_n^{(0)} + K\,b_n^{(1)} .
\]

Thus the system, to first order in $K$, becomes
\begin{center}
\begin{align}
\dot X &= \frac{\gamma}{2\pi}
\big[(b_1^{(0)} + K b_1^{(1)}) c_X - (a_1^{(0)} + K a_1^{(1)}) s_X\big] - \nu X,\\
\nonumber \dot a_0^{(0)} + K \dot a_0^{(1)} &= -\Big[(1 - 4\eta K) a_0^{(0)} + K a_0^{(1)} 
           - (1 - 2K)\frac{\eta}{\alpha} + 2K\alpha(c_X a_1^{(0)} + s_X b_1^{(0)})\Big],\\
\nonumber \dot a_1^{(0)} + K a_1^{(1)} &= -\Big[(1 - 4\eta K) a_1^{(0)} + K a_1^{(1)} 
           + (1 - 2K) c_X + 2K\alpha(2 c_X a_0^{(0)} + c_X a_2^{(0)} + s_X b_2^{(0)})\Big],\\
\nonumber \dot b_1^{(0)} + K \dot b_1^{(1)} &= -\Big[(1 - 4\eta K) b_1^{(0)} + K b_1^{(1)} 
           + (1 - 2K) s_X + 2K\alpha(2 s_X a_0^{(0)} + s_X a_2^{(0)} + c_X b_2^{(0)})\Big],\\
           n &\geq 2: \nonumber\\
\nonumber \dot a_n^{(0)} + K a_n^{(1)} &= -\Big[(1 - 4\eta K) a_n^{(0)} + K a_n^{(1)} 
           + 2K\alpha\big((a_{n+1}^{(0)} + a_{n-1}^{(0)}) c_X 
           + (b_{n+1}^{(0)} - b_{n-1}^{(0)}) s_X\big)\Big],\\
\nonumber \dot b_n^{(0)} + K \dot b_n^{(1)} &= -\Big[(1 - 4\eta K) b_n^{(0)} + K b_n^{(1)} 
           + 2K\alpha\big((a_{n+1}^{(0)} + a_{n-1}^{(0)}) s_X 
           + (b_{n+1}^{(0)} + b_{n-1}^{(0)}) c_X\big)\Big].
\end{align}
\end{center}

At zeroth order in $K$ we get
\begin{center}
\begin{align}
\dot a_0^{(0)} &= -\Big[a_0^{(0)} - \frac{\eta}{\alpha}\Big],\\
\nonumber \dot a_1^{(0)} &= -\Big[a_1^{(0)} + c_X\Big],\\
\nonumber \dot b_1^{(0)} &= -\Big[b_1^{(0)} + s_X\Big],\\
n &\geq 2: \nonumber\\
\nonumber \dot a_n^{(0)} &= -a_n^{(0)},\\
\nonumber \dot b_n^{(0)} &= -b_n^{(0)}.
\end{align}
\end{center}

From this system it is straightforward to see that $a_0^{(0)}$ decays exponentially to $\tfrac{\eta}{\alpha}$, so that $\alpha a_0^{(0)} = \eta$, representing the mean number of active motors when $K = 0$.  
The terms with $n \ge 2$ instead decay exponentially to zero.  
Thus the only non-trivial dynamics occur for $n = 1$.

The truly distinct phenomenology of the model with weak couplings emerges at first order in $ K $.  
In this case, we obtain:

\begin{center}
\begin{align}
\dot a_0^{(1)} &= -\Big[- 4\eta\, a_0^{(0)} + a_0^{(1)} + 2\frac{\eta}{\alpha} + 2\alpha(c_X a_1^{(0)} + s_X b_1^{(0)})\Big],\\
\nonumber \dot a_1^{(1)} &= -\Big[ - 4\eta\, a_1^{(0)} + a_1^{(1)} - 2 c_X + 2\alpha(2 c_X a_0^{(0)} + c_X a_2^{(0)} + s_X b_2^{(0)})\Big],\\
\nonumber \dot b_1^{(1)} &= -\Big[- 4\eta\, b_1^{(0)} + b_1^{(1)} - 2 s_X + 2\alpha(2 s_X a_0^{(0)} + s_X a_2^{(0)} + c_X b_2^{(0)})\Big],\\
n &\geq 2: \nonumber\\
\nonumber \dot a_n^{(1)} &= -\Big[-4\eta\, a_n^{(0)} + a_n^{(1)} + 2\alpha\big((a_{n+1}^{(0)} + a_{n-1}^{(0)}) c_X + (b_{n+1}^{(0)} - b_{n-1}^{(0)}) s_X\big)\Big],\\
\nonumber \dot b_n^{(1)} &= -\Big[-4\eta\, b_n^{(0)} + b_n^{(1)} + 2\alpha\big((a_{n+1}^{(0)} + a_{n-1}^{(0)}) s_X + (b_{n+1}^{(0)} + b_{n-1}^{(0)}) c_X\big)\Big].
\end{align}
\end{center}

This system is less trivial than the previous one and is coupled to zero-order terms.  
However, if we look at it for sufficiently long times to use the previous stationary values, namely  
$ a_0^{(0)} = \eta / \alpha $ and $ a_n^{(0)}, b_n^{(0)} = 0 $ for $ n \geq 2 $, then the system reduces to:

\begin{center}
\begin{align}
\dot a_0^{(1)} &= -\Big[-4\frac{\eta^2}{\alpha} + a_0^{(1)} + 2\frac{\eta}{\alpha} + 2\alpha(c_X a_1^{(0)} + s_X b_1^{(0)})\Big],\\
\nonumber \dot a_1^{(1)} &= -\Big[-4\eta\, a_1^{(0)} + a_1^{(1)} - 2 c_X + 4\eta c_X \Big],\\
\nonumber \dot b_1^{(1)} &= -\Big[-4\eta\, b_1^{(0)} + b_1^{(1)} - 2 s_X + 4\eta s_X \Big],\\
\nonumber \dot a_2^{(1)} &= -\Big[ a_2^{(1)} + 2\alpha\big(a_1^{(0)} c_X - b_1^{(0)} s_X\big)\Big],\\
\nonumber \dot b_2^{(1)} &= -\Big[ b_2^{(1)} + 2\alpha\big(a_1^{(0)} s_X + b_1^{(0)} c_X\big)\Big],\\
n &\geq 3: \nonumber\\
\nonumber \dot a_n^{(1)} &= -a_n^{(1)},\\
\nonumber \dot b_n^{(1)} &= -b_n^{(1)}.
\end{align}
\end{center}

For $ n \geq 3 $, the modes decay exponentially with unit characteristic time, they are therefore irrelevant in the long term, and it is reasonable to truncate the perturbative expansion at $ n = 2 $ without significant information loss.

The system that describes the problem after a sufficiently long time is thus:

\begin{center}
\begin{align}\label{eq:sistema pert
}
\dot X &= \frac{\gamma}{2\pi}\big[(b_1^{(0)} + K b_1^{(1)}) c_X - (a_1^{(0)} + K a_1^{(1)}) s_X\big] - \nu X,\\
\nonumber a_0^{(0)} &= \frac{\eta}{\alpha},\\
\nonumber \dot a_1^{(0)} &= -\big[a_1^{(0)} + c_X\big],\\
\nonumber \dot b_1^{(0)} &= -\big[b_1^{(0)} + s_X\big],\\
\nonumber a_n^{(0)} &= 0,\quad b_n^{(0)} = 0 \quad \text{for } n \geq 2,\\
\nonumber \dot a_0^{(1)} &= -\Big[-4\frac{\eta^2}{\alpha} + a_0^{(1)} + 2\frac{\eta}{\alpha} + 2\alpha(c_X a_1^{(0)} + s_X b_1^{(0)})\Big],\\
\nonumber \dot a_1^{(1)} &= -\Big[-4\eta\, a_1^{(0)} + a_1^{(1)} - 2 c_X + 4\eta c_X\Big],\\
\nonumber \dot b_1^{(1)} &= -\Big[-4\eta\, b_1^{(0)} + b_1^{(1)} - 2 s_X + 4\eta s_X\Big],\\
\nonumber \dot a_2^{(1)} &= -\Big[a_2^{(1)} + 2\alpha\big(a_1^{(0)} c_X - b_1^{(0)} s_X\big)\Big],\\
\nonumber \dot b_2^{(1)} &= -\Big[b_2^{(1)} + 2\alpha\big(a_1^{(0)} s_X + b_1^{(0)} c_X\big)\Big],\\
\nonumber a_n^{(1)} &= 0,\quad b_n^{(1)} = 0 \quad \text{for } n \geq 3.
\end{align}
\end{center}

The system exhibits a strongly hierarchical structure, in which the time derivatives of the first-order coefficients depend directly on the zeroth-order coefficients $ a_1^{(0)} $, $ b_1^{(0)} $, which act as forcing terms.  
No inverse coupling is present: the zeroth-order dynamics is unaffected by the first-order correction, in agreement with a perturbative approach.

All equations contain a negative linear relaxation term of the form $ \dot y = -y + \dots $, implying dissipative dynamics: in the absence of forcing, all components decay exponentially to zero.  
This suggests the existence of dynamical attractors (fixed points or limit cycles) to which the system converges in the long term.

It is interesting to note how moving from order 0 to order 1 makes the equations for higher $ n $ non-trivial (the truncation moves from $ n = 1 $ to $ n = 2 $).  
This is due to the structure of the unperturbed problem, which links nearest neighbours through the coupling constant $ K $, creating a link between $ a_n^{(1)}, b_n^{(1)} $ and $ a_{n-1}^{(0)}, b_{n-1}^{(0)} $.

One could investigate whether this feature, where at perturbative order $ K^m $, the terms up to $ a_{m+2}^{(m)}, b_{m+2}^{(m)} $ survive holds more generally.  
However, this makes no sense in our case, since a linear approximation was already made before expanding the coefficients in series.  
If this initial approximation could be avoided, this feature could potentially be explored.
In the simulations, we will use $\alpha=\eta=\frac{1}{2}$, so the equations can be further simplified by removing the dependence on these parameters, leading to the following easier system:

\begin{center}
\begin{align}\label{eq:sistema pert sim}
\dot X &= \frac{\gamma}{2\pi}\big[(b_1^{(0)} + K b_1^{(1)}) c_X - (a_1^{(0)} + K a_1^{(1)}) s_X\big] - \nu X,\\
\nonumber a_0^{(0)} &= 1,\\
\nonumber \dot a_1^{(0)} &= -\big[a_1^{(0)} + c_X\big],\\
\nonumber \dot b_1^{(0)} &= -\big[b_1^{(0)} + s_X\big],\\
\nonumber a_n^{(0)} &= 0,\quad b_n^{(0)} = 0 \quad \text{for } n \geq 2,\\
\nonumber \dot a_0^{(1)} &= -\Big[ a_0^{(1)}  + c_X a_1^{(0)} + s_X b_1^{(0)}\Big],\\
\nonumber \dot a_1^{(1)} &= -\Big[-2 a_1^{(0)} + a_1^{(1)} \Big],\\
\nonumber \dot b_1^{(1)} &= -\Big[-2 b_1^{(0)} + b_1^{(1)} \Big],\\
\nonumber \dot a_2^{(1)} &= -\Big[a_2^{(1)} + a_1^{(0)} c_X - b_1^{(0)} s_X\Big],\\
\nonumber \dot b_2^{(1)} &= -\Big[b_2^{(1)} + a_1^{(0)} s_X + b_1^{(0)} c_X\Big],\\
\nonumber a_n^{(1)} &= 0,\quad b_n^{(1)} = 0 \quad \text{for } n \geq 3.
\end{align}
\end{center}

{\color{red}\subsection{Numerical results for the decay of the coefficients}

\label{decay}

In this section, we complete the numerical evidence for the power-law decay of the Fourier coefficients as a function of the coupling parameter $K$, already introduced in Fig.~\ref{fig:Fourier} of the main text.
In particular, Fig.~\ref{fig:Fourier_complete} extends the analysis by including panels (f--h), which display the maxima of the Fourier coefficients $a_n$ and $b_n$ as functions of $K$ on a log--log scale, for $n = 1, \ldots, 6$.
In each panel, dashed lines represent power-law fits of the form $\sim K^{\alpha_a(n)}$ and $\sim K^{\alpha_b(n)}$, showing excellent agreement with the numerical data over several decades in $K$.
These results indicate that, starting from $n = 2$, the decay progressively approaches a clear power-law behaviour.

\begin{figure}[h!]
    \centering
        \includegraphics[width=0.99\textwidth]{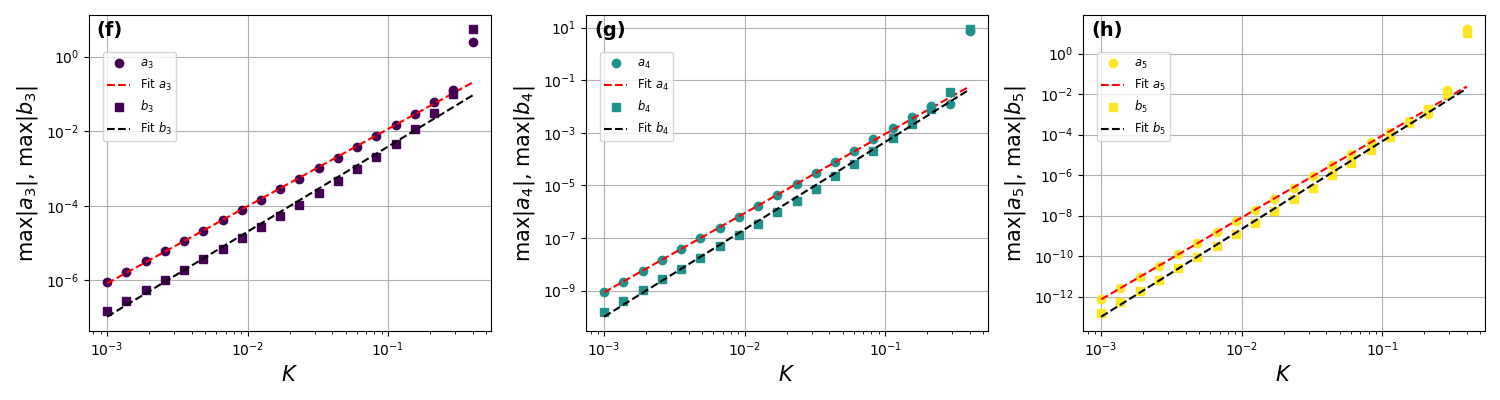}
    \caption{Numerical results about the decay of the maximum of the Fourier mode from the solution of the system \eqref{eq:sistema}.
    (f-h) maxima of the Fourier coefficients as functions of $K$; dashed lines indicate power-law fits $\sim K^{\alpha_a(n)}$ and $\sim K^{\alpha_b(n)}$.}
    \label{fig:Fourier_complete}
\end{figure}}

\subsection{In the flagellum frame}

We can switch to the reference frame of the flagellum, where the main results of the uncoupled model have been achived \cite{guerin2011bidirectional}, by changing variables as $\tilde \rho(x)=\rho(x+X)$.
For convenience,we can rename the density as $\rho$, we obtain the following drift terms:
\begin{align}
v[\rho] &= \frac{1}{\xi} \int_0^l dx \, W'(x) \rho(x) - \nu X, \\
\tilde{A}(x, \rho) &= \omega_{\text{off}}(x) \rho(x) - \omega_{\text{on}}(x) \left[ \frac{1}{l} - \rho(x) \right] - v[\rho] \frac{\partial \rho}{\partial x}, \\
\tilde{B}(x, \rho) &= \omega_{\text{off}}(x) \rho(x) + \omega_{\text{on}}(x) \left[ \frac{1}{l} - \rho(x) \right].
\end{align}

At this point, we can proceed with calculations exactly as in the reference frame where the motors are fixed. This leads to the following system for the evolution of the system:

\begin{center}
\begin{align}
     \dot X &= \frac{\gamma}{2 \pi} b - \nu X, \\
     \nonumber\dot a_0 &= -\left[ \Omega(1-4\eta K)a_0 - \frac{\Omega}{\ell}(1-2K)\frac{\eta}{\alpha} +2\Omega K \alpha a_1\right],\\
     \nonumber\dot a_1 &= -\left[\Omega(1-4\eta K)a_1+\frac{\Omega}{\ell}(1-2K)+2\Omega K \alpha(2a_0+a_2)-\frac{2\pi v}{l}b_1\right], \\
     \nonumber\dot b_1 &= -\left[\Omega(1-4\eta K)b_1+2\Omega K\alpha b_2+\frac{2\pi v}{l}a_1\right],\\
     \nonumber\dot a_n &= -\left[\Omega(1-4\eta K)a_n+2\Omega K \alpha(a_{n+1}+a_{n-1})-\frac{2\pi nv}{l}b_n\right],\\
     \nonumber\dot b_n &= -\left[\Omega(1-4\eta K)b_n+2\Omega K \alpha(b_{n+1}+b_{n-1})+\frac{2\pi nv}{l}a_n\right].
\end{align}
\end{center}

We observe that in this reference frame, the symmetry between $a_1$ and $b_1$ seen in the other frame is lost. The interactions between $a_n$ and $b_n$ only appear because of the velocity term $v$.

Moreover, except for the term proportional to the filament speed $v$, this system can be obtained from the one in the frame where the motors are fixed by setting $c_X = 1$ and $s_X = 0$, that is like assuming the filament is stationary for those terms.\\
Also in this case, we can proceed with a perturbative expansion as in the previous case. At order $K^0$, defining $\beta = 2\pi\nu$, and rescaling lengths and times so that $\ell = \Omega = 1$, we obtain:

\begin{center}
\begin{align*}
\dot a_0^{(0)} &= -\Big[a_0^{(0)} - \frac{\eta}{\alpha}\Big],\\
\dot a_1^{(0)} &= -\Big[a_1^{(0)} + 1 -\gamma\ (b_1^{(0)})^2+\beta b_1^{(0)}X \Big],\\
\dot b_1^{(0)} &= -\Big[b_1^{(0)}+\gamma\ b_1^{(0)}a_1^{(0)}-\beta a_1^{(0)}X\Big],\\
\dot a_n^{(0)} &= -[a_n^{(0)}+n\gamma b_1^{(0)}b_n^{(0)}-n \beta b_n^{(0)}X],\\
\dot b_n^{(0)} &= -[b_n^{(0)}-n\gamma b_1^{(0)} a_n^{(0)}+n \beta a_n^{(0)}X].
\end{align*}
\end{center}

This is exactly the same system that appears in the uncoupled model.\\
Also here, as in the other reference frame, the only non-trivial terms are $a_1^{(0)}$ and $b_1^{(0)}$, while $a_0^{(0)}$ tends exponentially to $\eta$ and $a_n^{(0)}, b_n^{(0)}$ tend exponentially to zero (the matrix that governs their evolution has trace -2 and determinant greater than 1).\\

At order $K^1$, the system becomes:

\begin{center}
\begin{align*}
\dot a_0^{(1)} &= -\Big[- 4\eta\, a_0^{(0)} + a_0^{(1)} + 2\frac{\eta}{\alpha} + 2\alpha a_1^{(0)}\Big],\\
\dot a_1^{(1)} &= -\Big[ - 4\eta\, a_1^{(0)} + a_1^{(1)} - 2 + 2\alpha(2a_0^{(0)} + a_2^{(0)}) - 2\gamma b_1^{(0)}b_1^{(1)} + \beta b_1^{(1)}X\Big],\\
\dot b_1^{(1)} &= -\Big[- 4\eta\, b_1^{(0)} + b_1^{(1)} - 2 + 2\alpha b_2^{(0)} + \gamma (b_1^{(0)}a_1^{(1)} + b_1^{(1)}a_1^{(0)}) - \beta a_1^{(1)}X\Big],\\
n &\geq 2: \nonumber\\
\dot a_n^{(1)} &= -\Big[-4\eta\, a_n^{(0)} + a_n^{(1)} + 2\alpha(a_{n+1}^{(0)} + a_{n-1}^{(0)}) - n\gamma (b_1^{(0)}b_n^{(1)} + b_1^{(1)}b_n^{(0)}) + n\beta b_n^{(1)}X\Big],\\
\dot b_n^{(1)} &= -\Big[-4\eta\, b_n^{(0)} + b_n^{(1)} + 2\alpha(b_{n+1}^{(0)} + b_{n-1}^{(0)}) + n\gamma (b_1^{(0)}a_n^{(1)} + b_1^{(1)}a_n^{(0)}) - n\beta a_n^{(1)}X\Big].
\end{align*}
\end{center}

Using the asymptotic values from the system at order 0, we can write:

\begin{center}
\begin{align*}
\dot a_0^{(1)} &= -\Big[- 4\frac{\eta^2}{\alpha} + a_0^{(1)} + 2\frac{\eta}{\alpha} + 2\alpha a_1^{(0)}\Big],\\
\dot a_1^{(1)} &= -\Big[ - 4\eta\, a_1^{(0)} + a_1^{(1)} - 2 + 4\eta - 2\gamma b_1^{(0)}b_1^{(1)} + \beta b_1^{(1)}X\Big],\\
\dot b_1^{(1)} &= -\Big[- 4\eta\, b_1^{(0)} + b_1^{(1)} - 2 + \gamma (b_1^{(0)}a_1^{(1)} + b_1^{(1)}a_1^{(0)}) - \beta a_1^{(1)}X\Big],\\
\dot a_2^{(1)} &= -\Big[a_2^{(1)} + 2\alpha a_1^{(0)} - 2\gamma b_1^{(0)}b_2^{(1)} + 2\beta b_2^{(1)}X\Big],\\
\dot b_2^{(1)} &= -\Big[b_2^{(1)} + 2\alpha b_1^{(0)} + 2\gamma b_1^{(0)}a_2^{(1)} - 2\beta a_2^{(1)}X\Big],\\
n &\geq 3: \nonumber\\
\dot a_n^{(1)} &= -\Big[a_n^{(1)} - n\gamma b_1^{(0)}b_n^{(1)} + n\beta b_n^{(1)}X\Big],\\
\dot b_n^{(1)} &= -\Big[b_n^{(1)} + n\gamma b_1^{(0)}a_n^{(1)} - n\beta a_n^{(1)}X\Big].
\end{align*}
\end{center}

For terms with $n \geq 3$, the matrix form is the same as the one found at order 0 for $n \geq 2$, so these terms decay exponentially to zero.\\
As in the previous reference frame, the correction at order $K$ forces us to keep more terms compared to the simple $K^0$ order, increasing the truncation limit from $n=1$ to $n=2$.\\

The system at order $K$ that describes the evolution of the problem is therefore:

\begin{center}
\begin{align}
\dot X &= \frac{\gamma}{2\pi}(b_1^{(0)} + K b_1^{(1)}) - \nu X, \\
\nonumber a_0^{(0)} &= \frac{\eta}{\alpha},\\
\nonumber \dot a_1^{(0)} &= -\Big[a_1^{(0)} + 1 - \gamma (b_1^{(0)})^2 + \beta b_1^{(0)}X\Big],\\
\nonumber \dot b_1^{(0)} &= -\Big[b_1^{(0)} + \gamma b_1^{(0)}a_1^{(0)} - \beta a_1^{(0)}X\Big],\\
\nonumber a_n^{(0)} &= b_n^{(0)} = 0 \quad \text{for} \quad n \geq 2,\\
\nonumber \dot a_0^{(1)} &= -\Big[- 4\frac{\eta^2}{\alpha} + a_0^{(1)} + 2\frac{\eta}{\alpha} + 2\alpha a_1^{(0)}\Big],\\
\nonumber \dot a_1^{(1)} &= -\Big[ - 4\eta a_1^{(0)} + a_1^{(1)} - 2 + 4\eta - 2\gamma b_1^{(0)}b_1^{(1)} + \beta b_1^{(1)}X\Big],\\
\nonumber \dot b_1^{(1)} &= -\Big[- 4\eta b_1^{(0)} + b_1^{(1)} - 2 + \gamma(b_1^{(0)}a_1^{(1)} + b_1^{(1)}a_1^{(0)}) - \beta a_1^{(1)}X\Big],\\
\nonumber \dot a_2^{(1)} &= -\Big[a_2^{(1)} + 2\alpha a_1^{(0)} - 2\gamma b_1^{(0)}b_2^{(1)} + 2\beta b_2^{(1)}X\Big],\\
\nonumber \dot b_2^{(1)} &= -\Big[b_2^{(1)} + 2\alpha b_1^{(0)} + 2\gamma b_1^{(0)}a_2^{(1)} - 2\beta a_2^{(1)}X\Big].
\end{align}
\end{center}

The fixed points of this system can be found.
It is easy to verify that one fixed point of the system is:
\begin{center}
\begin{align*}
X=0\,,\,\,a_0^{(0)}=\frac{\eta}{\alpha} \,\,,\,\,a_0^{(1)}=\frac{4\eta^2}{\alpha}-\frac{2\eta}{\alpha}+2\alpha \,\,,\,\, a_1^{(0)}=-1\\
a_1^{(1)} =2-8\eta \,\,,\,\, a_2^{(1)}=2\alpha \quad \text{and all the other terms vanish.}
\end{align*}
\end{center}

This leads to a motor density corrected at order $K$:
\begin{equation}
\rho(x)=\alpha\sum_n a_n c_n + b_n s_n = \eta + K(4\eta^2 - 2\eta + 2\alpha^2) + \left(-1 + 2K(1 - 4\eta)\right)\alpha c_1 + 2K\alpha^2 c_2.
\end{equation}

\subsection{The Role of Noise}

To analyse the noise, we work in the reference frame of the flagellum, following the same approach as in the uncoupled case \cite{Ma2014}, in order to compare the effects of coupling with the results obtained in the uncoupled system.\\

The noise term from the Fokker-Planck equation is given by:
\begin{equation}\label{eq:B}
    \tilde{B}(x, \rho) = \omega_{\text{off}}(x) \rho(x) + \omega_{\text{on}}(x) \left[ \frac{1}{\ell} - \rho(x) \right].
\end{equation}

Noting that $\omega_{\text{off}}(x) - \omega_{\text{on}}(x) = \Omega - 2\omega_{\text{on}}(x)$, we obtain:
\begin{equation*}
    \tilde{B}(x, \rho) = \Omega \rho(x) - 2\omega_{\text{on}}(x)\rho(x)\ell + \frac{\omega_{\text{on}}}{\ell}.
\end{equation*}

In the case with coupling-dependent rates, we can expand the exponential for small coupling constant $K$, as done previously:
\begin{equation}
    \omega_{\text{on}}(x) = \Omega \left[\eta - \cos\left(\frac{2\pi x}{\ell}\right)\right] e^{-2K(1 - 2\rho(x)\ell)} \simeq f(x)\left(1 - 2K(1 -  2\rho(x)\ell)\right).
\end{equation}

Given this, we can rewrite $\tilde{B}(x, \rho)$ in the following form, assuming for simplicity $\ell = 1$:
\begin{align*}
    \frac{\tilde{B}(x, \rho)}{\Omega} = (1 - 2\eta)\rho + 4K\eta\rho + 2\alpha c_1\rho - 4K\alpha c_1\rho - 8Kf\rho^2 + \eta(1 - 2K) - (1 - 2K)\alpha c_1 + 4Kf\rho.
\end{align*}

As we can see from this expression, in this case $\tilde{B}$ is not linear in $\rho$ but quadratic.\\
From an analytical point of view, developing this expression would require introducing all the coefficients from the expansion of $\rho$, which would make the equations formal but not very practical.\\
To obtain usable results, we instead expand $\rho(x)$ using the asymptotic values found in the deterministic case, corrected up to order $K$:
\begin{equation}
\rho(x) = \eta + K(4\eta^2 - 2\eta + 2\alpha^2) + \left(-1 + 2K(1 - 4\eta)\right)\alpha c_1 + 2K\alpha^2 c_2 = \eta + K\psi + \left(-1 + K\phi\right) c_1 + 2K\alpha^2 c_2.
\end{equation}

In this expression, $\psi = 4\eta^2 - 2\eta + 2\alpha^2$ and $\phi = 2\alpha - 8\alpha\eta$.\\
The equation further simplifies when considering terms like $K\rho(x)$, since some coefficients become of order $K^2$, and thus at order $K$ we obtain:
\begin{equation*}
    K\rho(x) = K(\eta - \alpha c_1).
\end{equation*}

We have:
\begin{align*}
    &K\rho(x)f(x) = K(\eta - \alpha c_1)^2 = K(\eta^2 - 2\alpha \eta c_1 + \alpha^2 c_1^2),\\
    &K\rho(x)f^2(x) = K(\eta - \alpha c_1)^3 = K(\eta^3 - 3\alpha \eta^2 c_1 + 3 \alpha^2 \eta c_1^2 - \alpha^3 c_1^3).
\end{align*}

From these expressions, we find:
\begin{align*}
    \frac{\tilde{B}(x,\rho)}{\Omega} &= (1 - 2\eta)(\eta + K\psi + (-1 + K\phi)c_1 + 2K\alpha^2 c_2) + 4K\eta(\eta - \alpha c_1)+ \\
    &\quad + 2\alpha c_1(\eta + K\psi + (-1 + K\phi)c_1 + 2K\alpha^2 c_2) - 4K\alpha c_1(\eta - \alpha c_1)+ \\
    &\quad - 8K(\eta^3 - 3\alpha \eta^2 c_1 + 3\alpha^2 \eta c_1^2 - \alpha^3 c_1^3) + \eta(1 - 2K)+ \\
    &\quad - (1 - 2K)\alpha c_1 + 4K(\eta^2 - 2\alpha \eta c_1 + \alpha^2 c_1^2)= \\
    &= 2(1 - \eta)\eta + K\left[(1 - 2\eta)\psi - 2\eta + 4\eta^2 - 8\eta^3\right]+ \\
    &\quad + c_1\left[-\alpha(1 - 2\alpha \eta) + K\left((1 - 2\eta)\phi - 4\alpha + \alpha\psi - 4\alpha\eta + 24\alpha \eta^2 + 2\alpha - 8\alpha \eta\right)\right]+ \\
    &\quad + c_1^2\left[-2\alpha^2 + K(2\alpha \phi + 8\alpha^2 - 24\alpha^2 \eta)\right]+ \\
    &\quad + c_1^3[8\alpha^3] + c_2[2K(1 - 2\eta)\alpha^2] + c_1 c_2[4\alpha^3].
\end{align*}

As in, we define the components of the diffusion matrix as:
\begin{align}
    D^{aa}_{mn} &= \frac{2}{N \alpha^2} \int_0^1 B(x)\, c_m\, c_n \, dx, \\
    D^{ab}_{mn} &= \frac{2}{N \alpha^2} \int_0^1 B(x)\, c_m\, s_n \, dx, \\
    D^{bb}_{mn} &= \frac{2}{N \alpha^2} \int_0^1 B(x)\, s_m\, s_n \, dx.
\end{align}

The only non-zero coefficients are $D_a = D^{aa}_{11}$ and $D_b = D^{bb}_{11}$.\\
The only relevant integrals for their computation are:
\begin{align*}
    \int_0^1 c_1^2 dx = \int_0^1 s_1^2 dx = \frac{1}{2}, \quad 
    \int_0^1 c_1^4 dx = \frac{3}{8}, \quad 
    \int_0^1 c_1^2 s_1^2 dx = \frac{1}{8}, \quad 
    \int_0^1 c_1^2 c_2 dx = \frac{1}{4}.
\end{align*}

We thus obtain the coefficients:
\begin{align*}
    D_a &= \frac{1}{2N}\left[\frac{4\eta(1 - \eta)}{\alpha^2} - 3\right] + \frac{K}{N\alpha^2} \left[(1 - 2\eta)\psi - 2\eta + 4\eta^2 - 8\eta^3 + (1 - 2\eta)\alpha^2 + \frac{3\phi \alpha}{2} + 6\alpha^2 - 18\eta \alpha^2 \right]. \\
    D_b &= \frac{1}{2N}\left[\frac{4\eta(1 - \eta)}{\alpha^2} - 1\right] + \frac{K}{N\alpha^2} \left[(1 - 2\eta)\psi - 2\eta + 4\eta^2 - 8\eta^3 - (1 - 2\eta)\alpha^2 + \frac{\phi \alpha}{2} + 2\alpha^2 - 6\eta \alpha^2 \right].
\end{align*}

By substituting the explicit expressions of $\phi$ and $\psi$ in terms of $\eta$ and $\alpha$, we obtain the final expressions for the coefficients:
\begin{align}
    D_a &= \frac{1}{2N}\left[\frac{4\eta(1 - \eta)}{\alpha^2} - 3\right] + \frac{K}{N\alpha^2} \left[10\alpha^2 + 2\alpha + 12 - 32\eta \alpha^2 - 4\eta \alpha - 16\eta^3-4\eta \right], \\
    D_b &= \frac{1}{2N}\left[\frac{4\eta(1 - \eta)}{\alpha^2} - 1\right] + \frac{K}{N\alpha^2} \left[2\alpha^2 + 2\alpha + 12\eta^2 - 8\eta \alpha^2 - 4\eta \alpha - 16\eta^3 -4\eta\right].
\end{align}

The structure of these coefficients is of the form $D_i = N^{-1} d_i^{(0)} + K N^{-1} d_i^{(1)}$.\\
The $d_i^{(0)}$ matches exactly the coefficients found in the model without coupling, so we can interpret the $d_i^{(1)}$ as the linear corrections due to weak coupling.\\

In that article, the diffusion coefficient $D_b$ is also related to the quality factor $Q$:
\begin{equation}\label{eq:Q K}
    Q \propto \frac{1}{D_b} = \frac{N}{d_b^{(0)} + K d_b^{(1)}}.
\end{equation}

In most of the simulations of this and previously cited articles, the values $\eta = \alpha = \frac{1}{2}$ are used. In this case, we find $d_b^{(1)} = -6$, so as $K$ increases, the quality factor should increase too, at least in the zone $K<<1$.\\
It is also interesting to note that this equation for $ Q $ provides another physical constraint on $ K $.\\
In fact, since $ Q $ must be $ \geq 0 $, the condition $ d_b^{(0)} + K d_b^{(1)} > 0 $ must hold wherever the formula is valid.\\

\subsection{Reduction of the system}
In this section, we present a series of heuristic considerations based on the full system of equations obtained via Fourier expansion of the motor density. The goal is to derive a simplified yet effective description of the system's behaviour by means of progressively more drastic reductions.

The full system for the Fourier coefficients $a_n(t)$, $b_n(t)$ and the position of the filament $X(t)$, linearized to first order in $K$, reads:
\begin{align}
\dot{X} &= \frac{\gamma}{2\pi} \left[ b_1(t)\cos(2\pi X(t)) - a_1(t)\sin(2\pi X(t)) \right] - \nu X(t) \label{eq:full_X},\\
\dot{a}_0 &= -\left[-q + q a_0(t) + K c_1(t) \right] \label{eq:full_a0},\\
\dot{a}_1 &= -\left[q a_1(t) + q\cos(2\pi X(t)) + K \left( 2\cos(2\pi X(t)) a_0(t) + c_2(t) \right) \right] \label{eq:full_a1},\\
\dot{b}_1 &= -\left[q b_1(t) + q\sin(2\pi X(t)) + K \left( 2\sin(2\pi X(t)) a_0(t) + s_2(t) \right) \right] \label{eq:full_b1}.
\end{align}

For $n \geq 2$, we have:
\begin{align}
\dot{a}_n &= -\left[q a_n(t) + K \tilde{a}_n(t) \cos(2\pi X(t)) + K \delta b_n(t) \sin(2\pi X(t)) \right], \\
\dot{b}_n &= -\left[q b_n(t) + K \tilde{a}_n(t) \sin(2\pi X(t)) + K \tilde{b}_n(t) \cos(2\pi X(t)) \right],
\end{align}

with the following auxiliary definitions:
\begin{align*}
q &= 1 - 2K, \\
c_n(t) &= \cos(2\pi X(t)) a_n(t) + \sin(2\pi X(t)) b_n(t), \\
s_n(t) &= \sin(2\pi X(t)) a_n(t) + \cos(2\pi X(t)) b_n(t), \\
\tilde{a}_n(t) &= a_{n+1}(t) + a_{n-1}(t), \\
\tilde{b}_n(t) &= b_{n+1}(t) + b_{n-1}(t), \\
\delta b_n(t) &= b_{n+1}(t) - b_{n-1}(t).
\end{align*}

This infinite hierarchy can be truncated at $n = 1$, which still yields good agreement with the full $N$-motor numerical simulations.
\begin{figure}[h]
    \centering
    \includegraphics[width=0.5\linewidth]{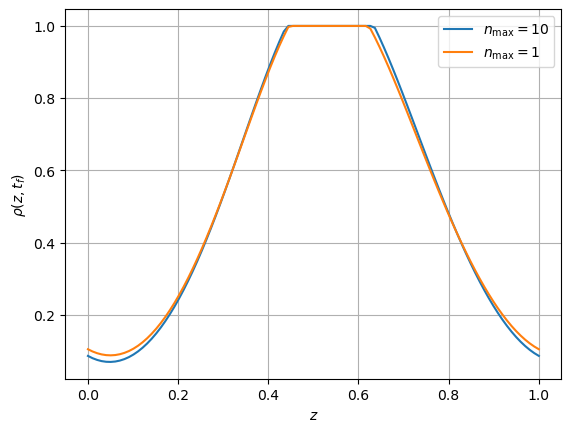}
    \caption{Effect of truncation on the system analysing the shape of density at $t_f=100$}
    \label{fig:tronc}
\end{figure}

We first consider a reduced system that retains only the coefficients $a_0$, $a_1$, and $b_1$:
\begin{align}
\dot{X} &= \frac{\gamma}{2\pi} \left[ b_1(t)\cos(2\pi X(t)) - a_1(t)\sin(2\pi X(t)) \right] - \nu X(t), \label{eq:reduced_X}\\
\dot{a}_0 &= -\left[-q + q a_0(t) + K c_1(t) \right], \label{eq:reduced_a0}\\
\dot{a}_1 &= -\left[q a_1(t) + q\cos(2\pi X(t)) + 2K a_0(t) \cos(2\pi X(t)) \right], \label{eq:reduced_a1}\\
\dot{b}_1 &= -\left[q b_1(t) + q\sin(2\pi X(t)) + 2K a_0(t) \sin(2\pi X(t)) \right], \label{eq:reduced_b1}
\end{align}

Now we can consider the time evolution of this coefficients.

\begin{figure}[h]
    \centering
    \includegraphics[width=0.7\linewidth]{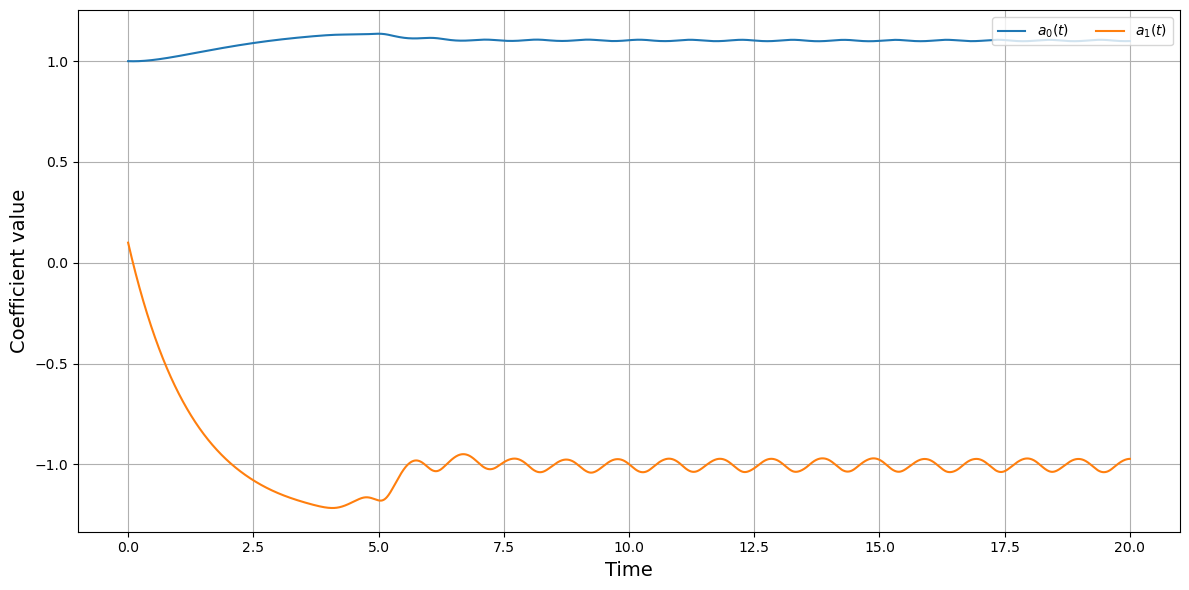}
    \includegraphics[width=0.7\linewidth]{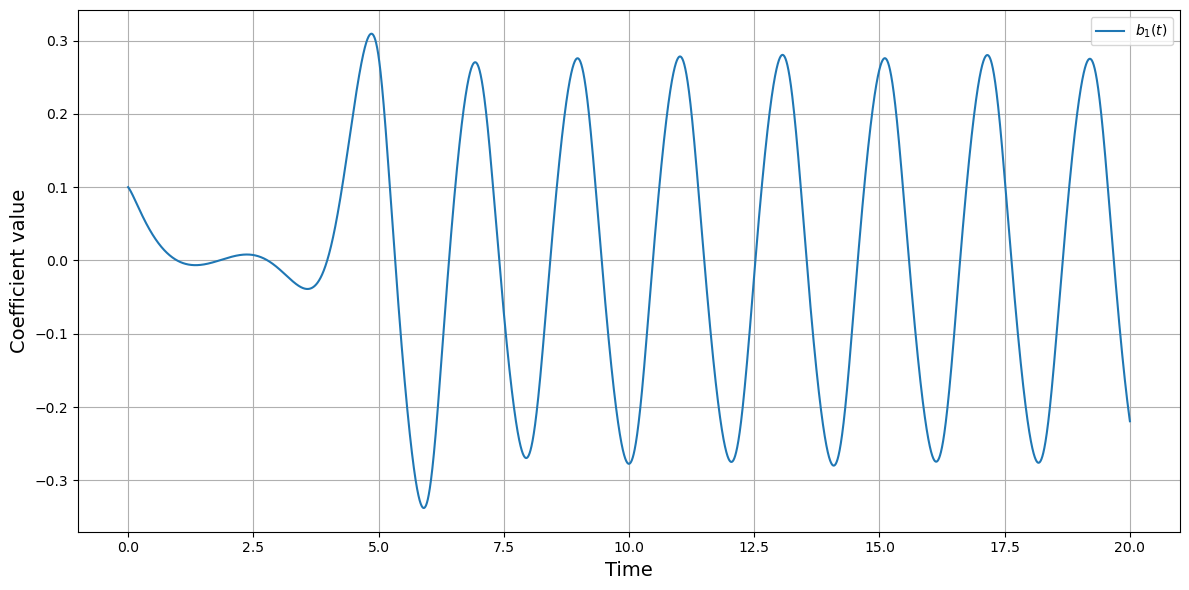}
    \caption{Evolution in time of the Fourier coefficients in case $K=0.1$.}
    \label{fig:abi}
\end{figure}

Since $a_0(t)$ quickly converges to a value close to 1, we can approximate:
\[
a_0(t) \approx 1,
\]
which leads to the simplified system:
\begin{align}
\dot{X} &= \frac{\gamma}{2\pi} \left[ b_1(t)\cos(2\pi X(t)) - a_1(t)\sin(2\pi X(t)) \right] - \nu X(t) \label{eq:approx_X},\\
\dot{a}_1 &= -\left[q a_1(t) + \cos(2\pi X(t)) \right] \label{eq:approx_a1},\\
\dot{b}_1 &= -\left[q b_1(t) + \sin(2\pi X(t)) \right] \label{eq:approx_b1}.
\end{align}

Similarly, $a_1(t)$ also tends to a nearly stationary value around $-1$. Therefore, we further approximate:
\[
a_1(t) \approx -1,
\]
and obtain a reduced system:

\begin{align}
\dot{X} &= \frac{\gamma}{2\pi} \left[ b_1(t)\cos(2\pi X(t)) + \sin(2\pi X(t)) \right] - \nu X(t) \label{eq:simplified_X},\\
\dot{b}_1 &= -\left[q b_1(t) + \sin(2\pi X(t)) \right] \label{eq:simplified_b1}.
\end{align}

Replacing the trigonometric expressions with their local linear approximations:
\begin{align*}
b_1(t) \cos(2\pi X(t)) &\to b_1(t), \\
\sin(2\pi X(t)) &\to 2\pi X(t).
\end{align*}
we obtain the minimal system:
\begin{align}
\dot{X} &= \frac{\gamma}{2\pi} \left[ b_1(t) + \sin(2\pi X(t)) \right] - \nu X(t) \label{eq:minimal_X},\\
\dot{b}_1 &= -\left[q b_1(t) + 2\pi X(t) \right]. \label{eq:minimal_b1}
\end{align}

Interestingly, replacing $\sin(2\pi X) \to X$ directly in Eq.~\eqref{eq:minimal_X} leads to instability in the dynamics even if the dissipative term dominates. This suggests that the sinusoidal term plays a crucial role in confining the system's trajectories.

The second equation admits a formal solution:
\[
b_1(t) = \int_{-\infty}^t e^{-q(t - s)} (-2\pi) X(s) \, ds.
\]
This expression shows that $b_1(t)$ acts as an exponentially weighted memory of the variable $X(t)$. As $q$ decreases, i.e., as $K$ increases, the memory becomes longer. In other words, increasing motor coupling leads to a stronger temporal integration of the filament position.

\end{appendix}


\bibliography{references}

\end{document}